\documentstyle[12pt,epsf,subeqn,axodraw,cite]{article}
\def\be{\begin{equation}}
\def\ee{\end{equation}}
\def\barr{\begin{array}}
\def\earr{\end{array}}
\def\dis{\displaystyle}
\def\ra{\rightarrow}
\def\gev{\; {\rm GeV} }
\def\bea{\begin{eqnarray}}
\def\eea{\end{eqnarray}}

\def\bold#1{\setbox0=\hbox{$#1$}
     \kern-.025em\copy0\kern-\wd0 
     \kern.05em\copy0\kern-\wd0 
     \kern-.025em\raise.0433em\box0 }

\def\cA#1{{\cal A}_{#1}}
\def\cB#1{{\cal B}_{#1}}
\def\cFg{{\cal F}_5^\gamma}
\def\cFZ{{\cal F}_5^Z}
\def\cg#1{{\cal H}_{#1}^\gamma}
\def\cZ#1{{\cal H}_{#1}^Z}
\def\ie{ {\em i.e.}}
\def\viz{ {\em viz.}}
\def\etal{ {\em et al.}}

\begin{document}
\vspace*{-1in}
\renewcommand{\thefootnote}{\fnsymbol{footnote}}
\begin{flushright}
CUPP--00/4 \\
\texttt{hep-ph/0011205} 
\end{flushright}
\vskip 5pt
\begin{center}
{\Large{\bf Trilinear Neutral Gauge Boson Couplings }}
\vskip 25pt
{\sf Debajyoti Choudhury $^{a,\!\!}$
\footnote{E-mail address: debchou@mri.ernet.in}}, 
{\sf Sukanta Dutta $^{b,\!\!}$
\footnote{E-mail address: Sukanta.Dutta@cern.ch}},   
{\sf Subhendu Rakshit $^{c,\!\!}$
\footnote{E-mail address: subhendu\_physics@yahoo.com}}  
and
{\sf Saurabh Rindani $^{d,\!\!}$
\footnote{E-mail address: saurabh@prl.ernet.in}}
\vskip 10pt 

$^a${\em Mehta Research Institute, Chhatnag Road, Jhusi, Allahabad
211 019, India }\\

$^b${\em Physics Department, S.G.T.B. Khalsa College, University of
Delhi, Delhi 110 007, India }\\

$^c${\em Department of Physics, University of Calcutta, 92 Acharya
Prafulla Chandra Road, \\ Calcutta 700 009, India} \\

$^d${\em Theory Group, Physical Research Laboratory, Navrangpura,
Ahmedabad 380 009, India }
\vskip 20pt

{\bf Abstract}
\end{center}

\noindent
\begin{quotation}
{\small We study the CP even trilinear neutral gauge boson vertices at
one-loop in the context of the Standard Model and the Minimal
Supersymmetric Standard Model, assuming two of the vector bosons are
on-shell. We also study the changes in the form-factors when these two
bosons are off-shell.
\vskip 10pt
\noindent   
} 
\end{quotation}

\vskip 20pt  

\setcounter{footnote}{0}
\renewcommand{\thefootnote}{\arabic{footnote}}
\section{Introduction}
        \label{sec:intro} 
Present measurements of the vector boson fermion couplings at the LEP
and the SLC remarkably confirm the Standard Model (SM) predictions to
a high degree of accuracy. While this strengthens our belief that the
weak interactions are indeed governed by a non-abelian gauge theory,
this hypothesis can be established only with an experimental
confirmation of the non-abelian structure of the SM.  Recently, some
progress has been made in this direction in experiments at
Tevatron~\cite{Teva} as well as at LEP-II~\cite{LEP2}.  However, the
relatively low sensitivity of such experiments does not allow us to
explore the couplings to the level of accuracy required to establish
the gauge-theoretic nature of the SM.  Nevertheless, one expects that
the significantly improved facilities available at future experiments
such as those at Linear Colliders (LC)~\cite{LC,CLIC,AFS}, would allow
us to corroborate the SM predictions in this sector.  Furthermore, an
accurate measurement of these cubic and quartic couplings could even
act as a pointer to the existence of new physics beyond the SM even at
energies lower than the corresponding production threshold.

Gauge invariance dictates that, within the SM, the trilinear neutral
gauge boson vertices (TNGBVs) vanish at the tree level. However,
one-loop corrections do generate small but non-vanishing values for
these couplings. In composite models, on the other hand, these
couplings can be significantly larger. In either case, one expects the
strength of these couplings to have a nontrivial dependence on the
momentum scale, a fact that may have a substantial bearing on their
experimental signature.

At LEP (Tevatron), the $ZZ\gamma$ and $Z\gamma\gamma$ vertices are
best studied in $Z\gamma$ production through processes such as
$e^+e^-\rightarrow Z\gamma$ ($q^+q^-\rightarrow Z\gamma$). The
anomalous coupling being of a non-renormalizable nature, a constant
value of the same would, in general, lead to a cross section growing
rapidly with energy. A momentum suppression, often expressed as a
form-factor behaviour, can ameliorate the non-unitary nature though.

LEP-II~\cite{LEP2} has been running at energies above $ZZ$ production
threshold and for the first time $Z$ pairs are being obtained. During
Run 2 at Tevatron and in future experiments at NLC or JLC, several
hundreds of such $Z$ pairs will be produced.  These could be
profitably used to constrain both $ZZZ$ and $\gamma ZZ$
vertices. These anomalous couplings manifest themselves differently in
the production of longitudinally or transversely polarised $Z$
bosons. Thus, for $ZZ$ production, helicity dependence of the decay
distributions constitute an additional source of information. In this
context it is worth mentioning that $WWV$ ($V=\gamma,\, Z$) vertex has
been analysed in detail within SM and as well as in supersymmetric
extension of it~\cite{bella}. The measurement of $WWZ$ and $WW\gamma$
couplings at LEP-II has deservedly received considerable attention.

Since any significant modification due to new interactions beyond the
SM constitutes a signal for new physics, some of these couplings have
already been studied in the literature.  However, very few of these
papers~\cite{baur,BBCD} treat the various qualitative and quantitative
issues in an adequate manner. Therefore there exists a good motivation
for a thorough and careful reexamination of the various contributions
within the SM as well as within one of its most popular
extension,\viz\ the Minimal Supersymmetric Standard Model (MSSM).
While this work was being completed, the papers of
Gounaris\etal~\cite{gounaris1,gounaris2} appeared and we will comment
on their results and make a comparison with our results later on.

In this paper we study the CP conserving couplings of TNGBVs. We
organize our paper as follows. In section 2 we describe a general
framework how the CP conserving form-factors can be derived from the
general tensorial structure of the three-point functions.  In section
3 we examine a general one-loop fermionic contribution to the three
point functions $\gamma^\star Z\gamma$, $Z^\star Z\gamma$,
$\gamma^\star ZZ$, $Z^\star Z Z$. In section 4 we calculate the SM
contribution to these couplings and in section 5 we extend our study
to MSSM. We study the changes in the form-factors in section 6 when
the all the gauge bosons are put off-shell. Section 7 contains our
conclusions.

\section{Generic structure of form-factors}
        \label{sec:generic} 
Let us consider the vertex: $V_1(p_{1\alpha})\, V_2(p_{2\beta})\,
V_3(p_{3\mu})$ where $V_i \equiv \gamma, Z$ and $p_1+p_2+p_3 = 0$.
The most general CP conserving tensorial structure for a three point
function can be written as\footnote{We adopt the convention
$\epsilon_{0123}=1$.}
\be
\barr{rcl}
\Gamma_{\alpha \beta \mu} (p_1,p_2;p_3) 
& = & \dis
\epsilon_{\alpha \beta \mu \eta} 
     \left(  \cA1 \, p_1^\eta
           + \cA2 \, p_2^\eta
     \right)
+ \epsilon_{\alpha \beta \rho \eta} \, p_1^\rho \,
p_2^\eta \, 
    \left( \cA3 \, p_{1\mu} + \cA4 \, p_{2\mu}
    \right)
\\[1.25ex]
& + & \dis
 \epsilon_{\alpha \mu \rho \eta} \, p_1^\rho \,
p_2^\eta \, 
    \left(\cA5 \, p_{1\beta} + \cA6 \, p_{2\beta}
    \right) 
+ \epsilon_{\beta \mu \rho \eta} \, p_1^\rho \,
p_2^\eta \, 
    \left(\cA7 \, p_{1\alpha} + \cA8 \, p_{2\alpha} 
    \right)
\earr
\ee
where $\cA{i}\equiv \cA{i}(p_1,p_2)$. Using Schouten's identity
(nonexistence of a totally antisymmetric fifth-rank tensor), we can
eliminate two of the above form-factors, say $\cA5$ and $\cA8$.
Furthermore, Bose symmetry ($p_1\leftrightarrow p_2$ , $\alpha
\leftrightarrow \beta $) relates the remaining form-factors
pairwise. Thus, we finally have
\be
\barr{rcl}
\Gamma_{\alpha \beta \mu}^{\tiny \rm Bose\ Sym.}
(p_1,p_2;p_3) 
 & = & \dis
\epsilon_{\alpha \beta \mu \eta} 
        (\cB1 p_1^{\eta} - \bar{\cB1} p_2^{\eta}) 
+ \epsilon_{\theta \mu \rho \eta} p_1^{\rho}
p_2^{\eta} 
    (\cB2 p_{2\beta} \delta^\theta_\alpha 
     - \bar{\cB2}\delta^\theta_\beta  p_{1\alpha} 
    )
 \\[1.25ex]
 & + & \dis 
\epsilon_{\alpha \beta \rho \eta}p_1^{\rho} p_2^{\eta}

(\cB3 p_{1\mu} + \bar{\cB3} p_{2\mu})
\earr
        \label{master}
\ee
where $\cB{i} \equiv \cB{i}(p_1, p_2)$ and $\bar \cB{i}\equiv
\cB{i}(p_2,p_1)$.  Note that the requirements of gauge invariance
and/or current conservation would further eliminate some of the
remaining free parameters.

\subsection{The $\gamma \gamma Z$ vertex}
Gauge invariance implies $\cB1 = - p_2^2 \cB2$ (and similarly, $\bar
\cB1 = - p_1^2 \bar \cB2$). Dropping terms proportional to the photon
momenta, we have, then
\[ 
\barr{rcl}
\dis
\Gamma_{\alpha \beta \mu}^{\gamma \gamma Z}
(p_1,p_2;p_3) 
& = & \dis - \epsilon_{\alpha \beta \mu \eta} 
              (\cB2 p_2^2 p_1^\eta - \bar \cB2 p_1^2 p_2^\eta) 
        \\[1.5ex]
  & +& \dis \frac{1}{2} 
        \epsilon_{\alpha \beta \rho \eta}p_1^{\rho}
p_2^{\eta} 
        \left[ (\cB3 - \bar \cB3) (p_1 - p_2)_\mu 
             - (\cB3 + \bar \cB3) p_{3\mu}
        \right]
\earr
\]
Thus, if the $Z$ and $\gamma (p_1)$ be on-shell, and the second photon
be off-shell, we can rewrite 
\subequations
\be 
\dis
\Gamma_{\alpha \beta \mu}^{\gamma \gamma^\ast Z}
(p_1,p_2;p_3) 
= i \cg3 \,\epsilon_{\beta \alpha \mu \eta} p_1^{\eta} 
           + i \frac{\cg4}{m_Z^2} \; 
           \epsilon_{\beta \alpha \rho \eta}  
                \, p_1^{\rho} \, p_2^{\eta} \,
p_{2\mu} 
     \label{g_gst_Z}
\ee
The form-factors $\cg{i}$ can be related to those of 
Ref.~\cite{Hagiwara} through 
\be
\barr{rclcl}
\dis \cg3 & \equiv & 
           \dis \frac{p_2^2}{m_Z^2} \: h_3^\gamma & =
& \dis 
           -i \cB2 p_2^2 \: = \: \dis i \cB1
    \\[2ex]
\dis \cg4 & \equiv & 
           \dis \frac{p_2^2}{m_Z^2} \: h_4^\gamma & =
& 
           i (\bar \cB3 - \cB3) m_Z^2
\earr
     \label{g_gst_Z_b}
\ee
\endsubequations
In eqn.(\ref{g_gst_Z}), we have dropped the term proportional to
$p_{3\mu}$ assuming that the $Z$ couples to light fermions.

What if both photons are on-shell? Clearly, $\cB{i} = \bar \cB{i}$ in
this case, and
\[ \dis 
\Gamma_{\alpha \beta \mu}^{\gamma \gamma Z^\ast}
(p_1,p_2;p_3)
    = - \cB3 \epsilon_{\alpha \beta \rho \eta} 
                                p_1^{\rho} p_2^{\eta}
p_{3\mu}
\]
reflecting the well-known result that a massive vector boson cannot
decay into two massless vector bosons. The above can also be seen from 
eqns.(\ref{g_gst_Z_b}) whereby both of $\cg{3,4}$ vanish identically. 

\subsection{The $ZZ \gamma$ vertex}

The gauge invariance condition now reads 
$p_3^\mu \Gamma_{\alpha \beta \mu}^{ZZ\gamma} = 0 $,
leading to
\be
 \cB1 + \bar \cB1 - (p_1^2 + p_1 \cdot p_2) \cB3 
                - (p_2^2 + p_1 \cdot p_2) \bar \cB3 
  = 0
        \label{ZZg_gaugeinv}
\ee
When the photon is on-shell, this reduces to
\[
\cB1 + \bar \cB1 
     = (p_1^2 + p_1 \cdot p_2) (\cB3 - \bar \cB3) \ .
\]
Defining $\cB{1A} = (\cB1 - \bar \cB1) / 2$, and similarly for
$\cB{3A}$, we then have
\[  \dis
\Gamma_{\alpha \beta \mu}^{Z Z^\ast
\gamma}(p_1,p_2;p_3)
        = -\cB{1A} \epsilon_{\alpha \beta \mu \eta}
p_3^\eta
        + \cB{3A} \epsilon_{\alpha \beta \rho \eta}
                 \left[ p_1^\rho p_2^\eta (p_1 -
p_2)_\mu 
                        + (p_1^2 + p_1 \cdot p_2) (p_1
- p_2)^\eta 
                                   \delta^\rho_\mu
                 \right]
\]
which can be recast (on using Schouten's Identity) as
\[  \dis
\Gamma_{\alpha \beta \mu}^{Z Z^\ast
\gamma}(p_1,p_2;p_3)
        = - \epsilon_{\alpha \beta \mu \eta} p_3^\eta
\left[\cB{1A}
        + 2 \cB{1S} \frac{p_1^2+p_2^2}{p_1^2-p_2^2}
\right]
        + \cB{3A} \epsilon_{\alpha \mu \rho \eta}
p_1^\rho p_2^\eta p_{1\beta} 
        + \cB{3A} \epsilon_{\beta \mu \rho \eta}
p_1^\rho p_2^\eta p_{2\alpha}. 
\]
A rearrangement of terms then leads to 
\subequations
\be \dis
\Gamma_{\alpha \beta \mu}^{Z Z^\ast \gamma}
  = i \cZ3 \, \epsilon_{\beta \alpha \mu \eta} \,
p_3^{\eta} 
           + i \frac{\cZ4}{ m_Z^2} \: 
                \left[ p_{2\alpha} \,
                          \epsilon_{\beta \mu \rho \eta} \,
                        p_2^{\rho} \, p_3^{\eta} 
           +  p_{1\beta} \, \epsilon_{\alpha \mu  \rho \eta} \,
                        p_1^{\rho} \, p_2^{\eta} 
                \right]
       \label{H3Z}
\ee
where
\be
\barr{rclcl}
\cZ3 & \equiv & \dis \frac{m_Z^2 - p_2^2}{m_Z^2} \,
h_3^Z
       & = & \dis -i \left( \cB{1A} + 2 \cB{1S} \:
        \frac{p_1^2+p_2^2}{p_1^2-p_2^2} \right)
 \\[2ex]
\cZ4 & \equiv & \dis \frac{m_Z^2 - p_2^2}{m_Z^2} \,
h_4^Z
       & = & \dis -i m_Z^2 \cB{3A}
\earr
       \label{H3Z_b}
\ee
\endsubequations
with $\cB{1S}$ defined as $\cB{1S} = (\cB1 + \bar \cB1) / 2$.  It
should be noted that in eqn.~\ref{H3Z}, we have an extra form-factor
as compared to Ref.~\cite{Hagiwara}. However, since $\cZ4$ turns out
to be identically zero, one need not worry on this score.

On the other hand, when the photon is off-shell and both the $Z$'s are
on-shell (\ie, $\cB{i} = \bar \cB{i}$), the gauge condition simplifies
to $2 \cB1 = p_3^2 \cB3$ and
\subequations
\be 
\dis
\Gamma_{\alpha \beta \mu}^{Z Z \gamma^\ast} 
   = i \cFg
            \; \epsilon_{\alpha \beta \mu \eta} (p_1 -
p_2)^\eta
        \label{f5g_defn}
\ee
with
\be \dis
\cFg \equiv -\frac{p_3^2}{m_Z^2} \, f_5^{ZZ\gamma} 
         = -i \, \cB1.
        \label{f5g_b}
\ee
\endsubequations

\subsection{The $ZZZ$ vertex}
To derive the most general form, we need to impose the additional
symmetry ($p_1 \leftrightarrow p_3, \alpha \leftrightarrow \mu$) on
eqn.(\ref{master}). This, however, is not very illuminating.  Rather,
note that for real $Z$-pair production, again $\cB{i} = \bar \cB{i}$
and thus
\subequations
\be \dis
\Gamma_{\alpha \beta \mu}^{Z Z Z^\ast} 
   = i \cFZ
            \; \epsilon_{\alpha \beta \mu \eta} (p_1 -
p_2)^\eta
        \label{f5Z_defn}
\ee
with 
\be \dis
\cFZ \equiv -\frac{p_3^2 - m_Z^2}{m_Z^2} f_5^{ZZZ} 
     = -i \, \cB1. 
        \label{f5Z_b}
\ee
\endsubequations
For notational convenience we shall use $f_5^{ZZ\gamma}=f_5^{\gamma}$
and $f_5^{ZZZ}=f_5^{Z}$.  Explicit calculation within a model would
always ensure the proper symmetry structure.

\section{One-loop contributions to the TNGBVs}
        \label{sec:one-loop} 
Let us briefly examine the CP violating form-factors first.  For these
to exist at the one-loop level, one obviously needs the internal
states to have CP non-conserving couplings to the $Z$. The SM
particles, whether fermions or the Higgs, clearly do not meet the
requirement. That the $Z\gamma \gamma$ and $Z Z \gamma$ vertices will
continue to preserve CP even within the MSSM is also easy to see. The
$ZZZ$ vertex, on the other hand, can violate CP even at one-loop, but
only if CP non-conservation is introduced in the scalar sector. We
shall not consider this possibility here.

\begin{figure}[h]
\begin{center}
\begin{picture}(400,120)(0,0)
\SetWidth{1.2}
\Photon(80,50)(160,50){4}{8}
\Photon(220,100)(300,100){4}{8}
\Photon(220,0)(300,0){4}{8}
\ArrowLine(220,100)(160,50)
\ArrowLine(220,0)(220,100)
\ArrowLine(160,50)(220,0)
\Text(60,50)[c]{{$V_3^\mu$}}
\Text(320,100)[c]{{$V_1^\alpha$}}
\Text(320,0)[c]{{$V_2^\beta$}}
\LongArrow(100,70)(120,70)
\LongArrow(280,120)(260,120)
\LongArrow(280,-20)(260,-20)
\Text(90,70)[c]{{$p_3$}}
\Text(290,120)[c]{{$p_1$}}
\Text(290,-20)[c]{{$p_2$}}
\Text(187,88)[c]{{$f_a$}}
\Text(187,13)[c]{{$f_c$}}
\Text(230,50)[c]{{$f_b$}}
\end{picture}
\end{center}
\vskip 20pt
\caption[]{\em Generic diagram contributing to trilinear neutral gauge
boson vertices for the CP conserving case.}\label{fig:feyn}
\end{figure}
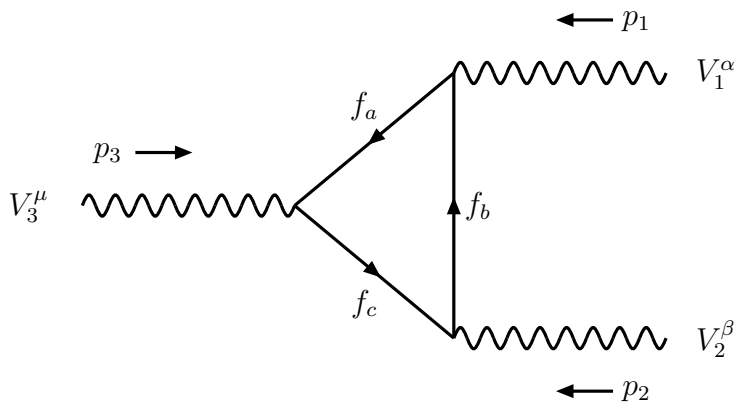

As for the CP conserving ones, again, to one-loop order, only the
fermions in the theory may contribute~\cite{Renard}. In
Fig.~\ref{fig:feyn}, we draw a generic diagram
\noindent
contributing to this process. Denoting the fermion-gauge coupling by
\[
    i e \bar f_a \gamma_\mu \left(l_i^{ab} P_L +
r_i^{ab} P_R
                            \right) \: f_b 
          \;  V_i^\mu           \ ,
\] 
with $P_{L, R} = (1 \mp \gamma_5) / 2$, it is useful to define the
combinations \be \barr{rcl} \varphi_1 & = & \dis l_1^{ab} l_2^{bc}
l_3^{ca} - r_1^{ab} r_2^{bc} r_3^{ca} \\[2ex] \varphi_2 & = & \dis m_a
m_c \left( l_1^{ab} l_2^{bc} r_3^{ca} - r_1^{ab} r_2^{bc} l_3^{ca}
\right) \\[2ex] \varphi_3 & = & \dis m_a m_b \left( l_1^{ab} r_2^{bc}
r_3^{ca} - r_1^{ab} l_2^{bc} l_3^{ca} \right) \\[2ex] \varphi_4 & = &
\dis m_b m_c \left( l_1^{ab} r_2^{bc} l_3^{ca} - r_1^{ab} l_2^{bc}
r_3^{ca} \right) \ .  \earr \label{the_combins} \ee Here $m_i$,
$i=a,b,c$, are the masses of the internal fermions $f_i$.  The
contributions of the diagram of Fig.~\ref{fig:feyn} can then be
parametrized as
\be
\barr{rcl}
(4 \pi/\alpha) \: \cB1 & = & \dis 
   \varphi_1 \Big[ p_3^{2}(C_{11}+C_{21}) -
p_2^{2}(C_{12} + C_{22})
            -  2 p_1\cdot p_2 (C_{12}+C_{23}) 
               \Big]  
   \\[1.5ex]
 &+ & \dis
     (m_a^2 \varphi_1 + \varphi_3 - \varphi_2)
(C_{11}+C_{0}) 
               - \varphi_4 C_{11}
 \\[2ex]
(4 \pi/\alpha) \: \bar{\cB1} & = & \dis
   - (m_a^2 \varphi_1 - \varphi_4 -
\varphi_2) (C_{11}-C_{12}) 
             - \varphi_3 (C_{11}+C_{0}-C_{12})
  \\[1.5ex]  
        & - & \dis 
\varphi_1 \Big[B_{023}+ B_{123} -2 C_{24} 
                + p_1^{2} (C_{12}+C_{21} )
  \\[1.5ex]  
        &  & \dis  \hspace*{4em}
                + (p_2^{2} + 2 p_1 \cdot p_2)
(C_{22}-2 C_{23}+C_{21}) \Big]
 \\[2ex]
(4 \pi/\alpha) \: \cB2 
        & = & \dis 2 \varphi_1 (C_{22}-C_{23})
 \\[2ex]
(4 \pi/\alpha) \: \bar{\cB2} 
         & = & \dis 2 \varphi_1 (C_{12}+C_{23})
 \\ [2ex]
(4 \pi/\alpha) \: \cB3 
        & = & \dis 2 \varphi_1
(C_{21}+C_{11}-C_{23}-C_{12})
 \\[2ex]
\bar{\cB3} & = &  \cB3  \ .
\earr \label{eq:loop}
\ee       

In eqn.(\ref{eq:loop}), the quantities $B$'s and $C$'s are the usual
Passarino-Veltman functions~\cite{veltman} relevant to the diagram in
question. We follow the following convention for the $C$ functions:
\bea 
C_{0;\,\mu;\,\mu\nu} & \equiv & \dis 
C_{0;\,\mu;\,\mu\nu}(p_3,p_2,m_a,m_c,m_b)\\\nonumber 
 &=& \dis
\frac{1}{i\pi^2} \int d^{4}k
\frac{1;\;k_{\mu};\;k_{\mu}k_{\nu}}
{(k^2+m_a^2)[(k+p_3)^2+m_c^2][(k+p_3+p_2)^2+m_b^2]} 
\eea
$B_{023}$ and $B_{123}$ denote $B_0(2,3)$ and $B_1(2,3)$
respectively. For all these conventions we follow Ref.~\cite{veltman}.
We have evaluated these Passarino-Veltman functions numerically using
existing numerical packages~\cite{ARCFF}.  An alternative method
involves calculation of the absorptive parts explicitly in terms of
simple integrals and then reexpressing the real parts in terms of
dispersion relations~\cite{SDR_PP}. We have checked that the two
methods give identical results.

To obtain the full contribution, one needs to consider all the
topologically distinct diagrams for a given set of fermions and, then,
add the contributions due to different sets.  It is clear that the
form-factors are ultraviolet finite.  They would be identically zero
in the limit of degenerate fermions which becomes apparent after we
sum over the fermions.

It is curious to note that $\bar{\cB3} = \cB3$ irrespective of the
fermion content, and hence $h_4^\gamma = 0 = h_4^Z$ to one-loop order.

\section{The SM contribution revisited}
        \label{sec:SM}

\begin{figure}[htb]
\hspace*{-0.5cm}
\centerline{
\epsfxsize=8.cm\epsfysize=8.0cm
\epsfbox{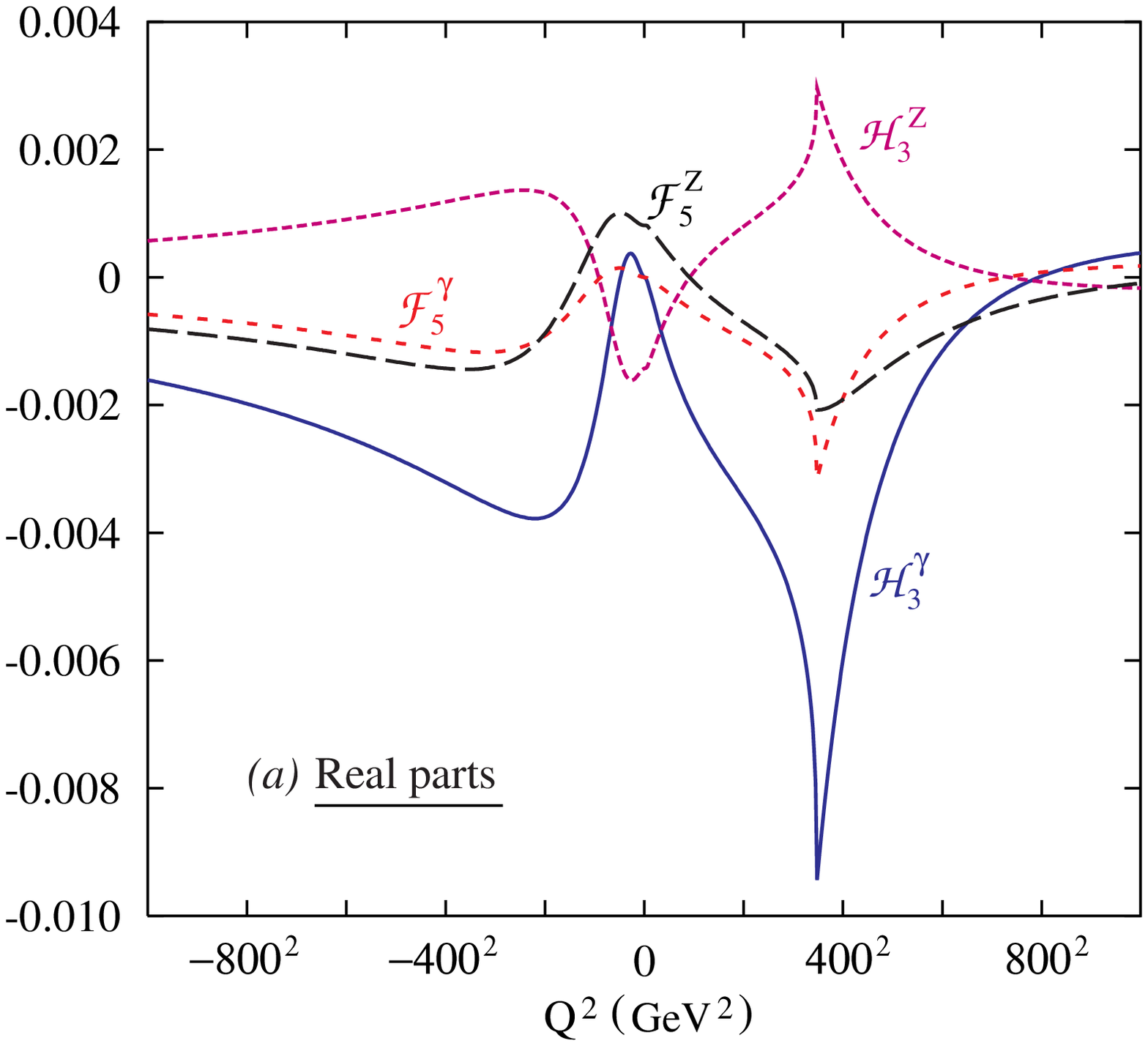} 
\vspace*{-0.0cm}
\hspace*{-0.5cm}
\epsfxsize=8.cm\epsfysize=8.0cm
\epsfbox{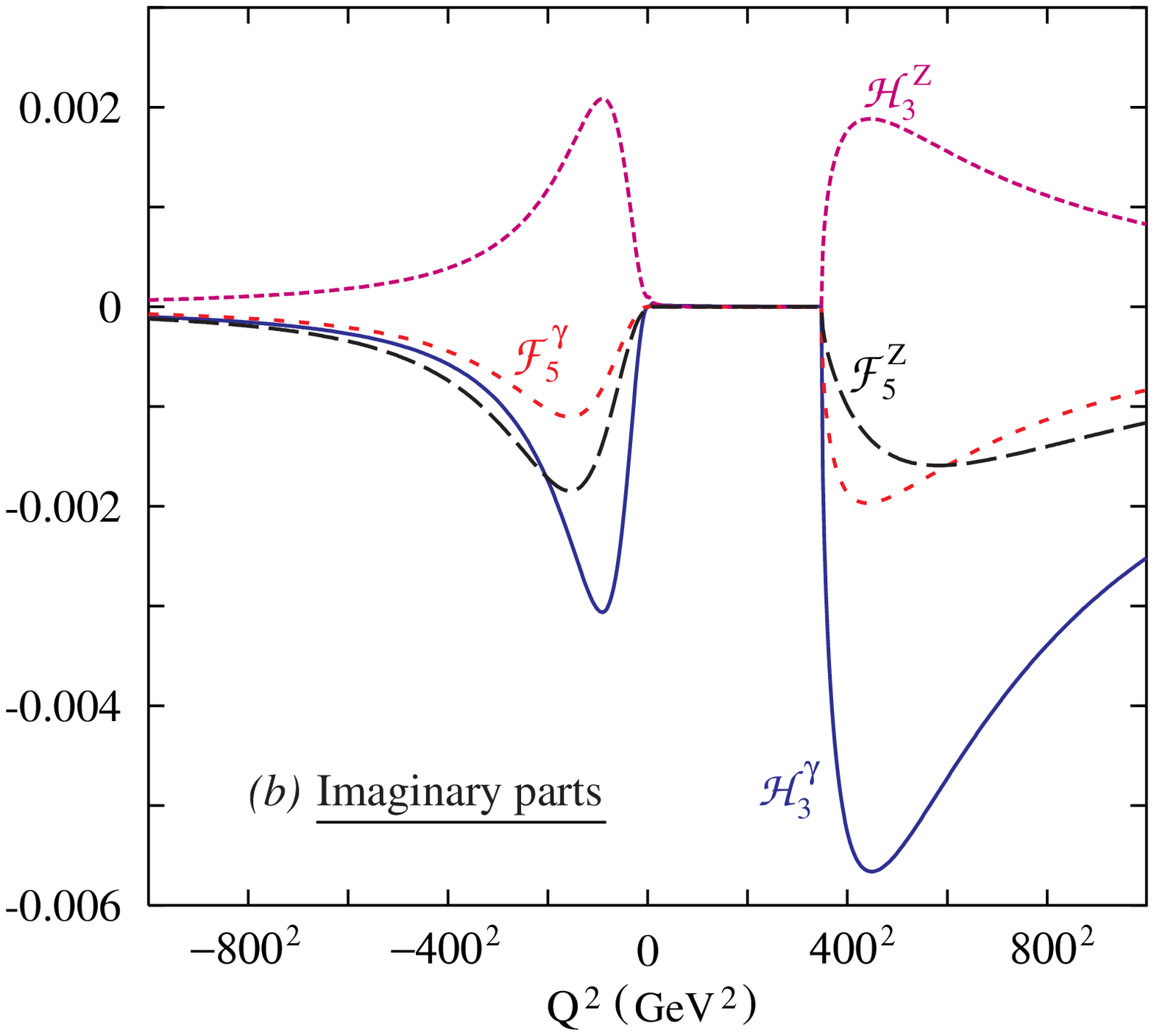}
\vspace*{-0.3cm}
}
\caption{\em The $Q^2$ dependence of the four non-zero
             form-factors within the SM ($m_t = 175 \gev$).}
      \label{fig:sm}
\end{figure}

Within the SM, all the couplings under consideration are identically
zero at the tree level.  However, at the one-loop level, charged
fermion loops contribute to the CP conserving form-factors.  $\cFZ$,
in addition, also receives contributions from the neutrino loops.
Initially, we restrict ourselves to the case where only one of the
three vector bosons is off-shell.  Denoting this momentum
transfer\footnote{Here $Q$ can be either time-like or space-like,
depending on whether the process is a $s$- or a $t$-channel one.}  by
$Q$, we present, in Figs.\ref{fig:sm}($a,b$), the real and imaginary
parts of the various form-factors as a function of $Q^2$.
%
%
For a very large $Q^2$, all the SM fermions behave as if they are
massless. Thus anomaly cancellation assures that asymptotically, all
these form-factors must vanish. For relatively smaller values of
$Q^2$, an analysis of eqns.(\ref{eq:loop}) shows that the relative
contribution of each fermion loop grows with the fermion mass.  The
maximum contribution, thus, occurs for the heaviest fermion.  On the
other hand, it is clear that while ${\cal H}^\gamma_3$ and ${\cal
F}_5^\gamma$ must vanish for $Q^2 = 0$, the other two (${\cal H}^Z_3$
and ${\cal F}_5^Z$) vanish as $Q^2 \ra m_Z^2$.

The imaginary parts of the four form-factors receive a contribution
from a particular fermion loop only when the kinematics allows two of
the fermions to be on-shell. This can happen for two different cases
\begin{enumerate} 
\item if $Q^2 > 4 m_f^2$ when the $s$-channel boson
goes to a real pair of fermions. In Fig.~\ref{fig:sm}($b$), this is
evinced by the top thresholds.  
\item for space-like momentum
transfers with a real $Z$ in the final state such that $m_Z > 2 m_f$
where $m_f$ is the mass of the fermion in the loop.  Clearly, the
top-quark can never contribute to the imaginary parts for such
``$t$-channel'' ({\it i.e.} $Q^2 < 0$) processes.  Since the light
fermion contributions essentially cancel amongst each other, the
magnitude of such imaginary parts are determined primarily by the $b$-
and $\tau$-loops.  
\end{enumerate}

For the real parts, the situation is analogous, but slightly more
complex. This part can be better in terms of a dispersion integral of
the absorptive part~\cite{SDR_PP}. The opening up of a channel now
manifests itself as a kink, rather than a typical threshold
jump. This, for example, is quite akin to the behaviour one sees in
Higgs production through gluon-gluon fusion. For $Q^2 < 0$, one does
not expect a threshold behaviour. That the form-factors have to fall
off as $Q^2 \to - \infty$ is obvious. The maximum shown by each curve
can intuitively be understood in terms of `phase-space available' as
in a $t$-channel scattering.


\section{TNGBVs within the MSSM}
        \label{sec:mssm} 
As we have already argued, to one-loop order, only the fermionic
sector of a model may contribute to the form-factors under
discussion. Going from the SM to the MSSM, the only augmentation of
the fermionic spectrum is in the form of the chargino-neutralino
sector. To recapitulate, the ($4 \times 4$) mass matrix for the
neutralinos is determined by four parameters, $M_{1,2}$, the soft
supersymmetry breaking mass parameters for the $U(1)$ and $SU(2)$
gauginos\footnote{We shall be assuming gaugino mass unification
subsequently.  Then $M_1$ and $M_2$ will be related as
$M_1=\frac{5}{3}\tan^2 \theta_W M_2$ and the neutralino mass matrix
will be determined by three independent parameters $M_2$, $\mu$ and
$\tan\beta$.}, the Higgsino mass parameter $\mu$ and $\tan \beta$, the
ratio of the vacuum expectation values of the two Higgs fields. The
orthogonal matrix $N$ that diagonalizes this (real) symmetric mass
matrix expresses the physical states in terms of the gauge eigenstates
and thus enters the interaction vertices.  On the other hand, the
chargino mass matrix (determined by $M_2$, $\mu$ and $\tan \beta$)
being real but nonsymmetric cannot be diagonalized by a single
orthogonal matrix. Rather, one needs two such matrices $U$ and $V$
that left- and right-diagonalize it respectively. Writing the
neutralino mass matrix in the $(\widetilde B, \widetilde W_3,
\widetilde H_1, \widetilde H_2)$ basis and the chargino mass matrix in
the $(\widetilde W^+, \widetilde H^+)$ basis, one can
express~\cite{HaberKane} the relevant electromagnetic and weak
currents as
\subequations
\be
{\cal J}^\mu_{e.m}=\sum_i \overline{\tilde\chi_i^+} \;\gamma^\mu
\tilde \chi_i^+
\ee
and
\be
{\cal J}^\mu_Z=\sum_{i ,j} \overline{\tilde\chi_i^+}\;\gamma^\mu
\left(P_L\,O'^L_{ij} +P_R\, O'^R_{ij}\right) \tilde
\chi_j^+
+
\sum_{\alpha ,\beta} \, O''_{\alpha\beta} \, 
        \overline{\tilde\chi^0_\alpha} \gamma^\mu \gamma_5
        \tilde\chi_\beta^0
\ee
where 
\be
\barr{rcl}
O'^L_{ij}& = & \dis  \sin^2\theta_W\,\delta_{ij} - 
                \left({1\over 2} V_{i2}V_{j2}^\star
                        + V_{i1}V_{j1}^\star\right) 
        \\[1.5ex]
O'^R_{ij} & = & \dis  \sin^2\theta_W\,\delta_{ij} - 
          \left({1\over 2}U_{i2}^\star U_{j2}\;
                 + U_{i1}^\star U_{j1}\right) 
        \\[1.5ex]
O''_{\alpha\beta}& = & \dis {1\over 4} \left(
 N_{\alpha 3}N_{\beta 3}^\star-N_{\alpha 4}N_{\beta 4}^\star \right)
\earr
        \label{the_O's}
\ee
\endsubequations

Armed with the above, we can now calculate the MSSM contributions to
the form-factors which were studied in the section~\ref{sec:SM} in the
context of the SM.

\subsection{Contribution to TNGBVs}
As in the case of the SM, we start with the $Q^2$ dependence of the
different form-factors. For this purpose, we choose a particular point
in the MSSM parameter space namely $M_2=100\gev$, $\mu=500\gev$ and
$\tan\beta=2$.  For this set of parameters the chargino masses turn
out to be $87.0\gev$ and $515.1\gev$, while the neutralino masses are
$45.9\gev$, $88.2\gev$, $501.4\gev$ and $517.7\gev$.  In each case,
the gaugino component is the predominant one as far as the lighter
eigenstates are concerned.  While the charginos contribute to all of
the TNGBVs, the neutralinos make their presence felt only in the
$ZZZ$ vertex.

\begin{figure}[htb]
\vspace*{-2.0cm}
\centerline{
\epsfxsize=7.0cm\epsfysize=9.0cm
\epsfbox{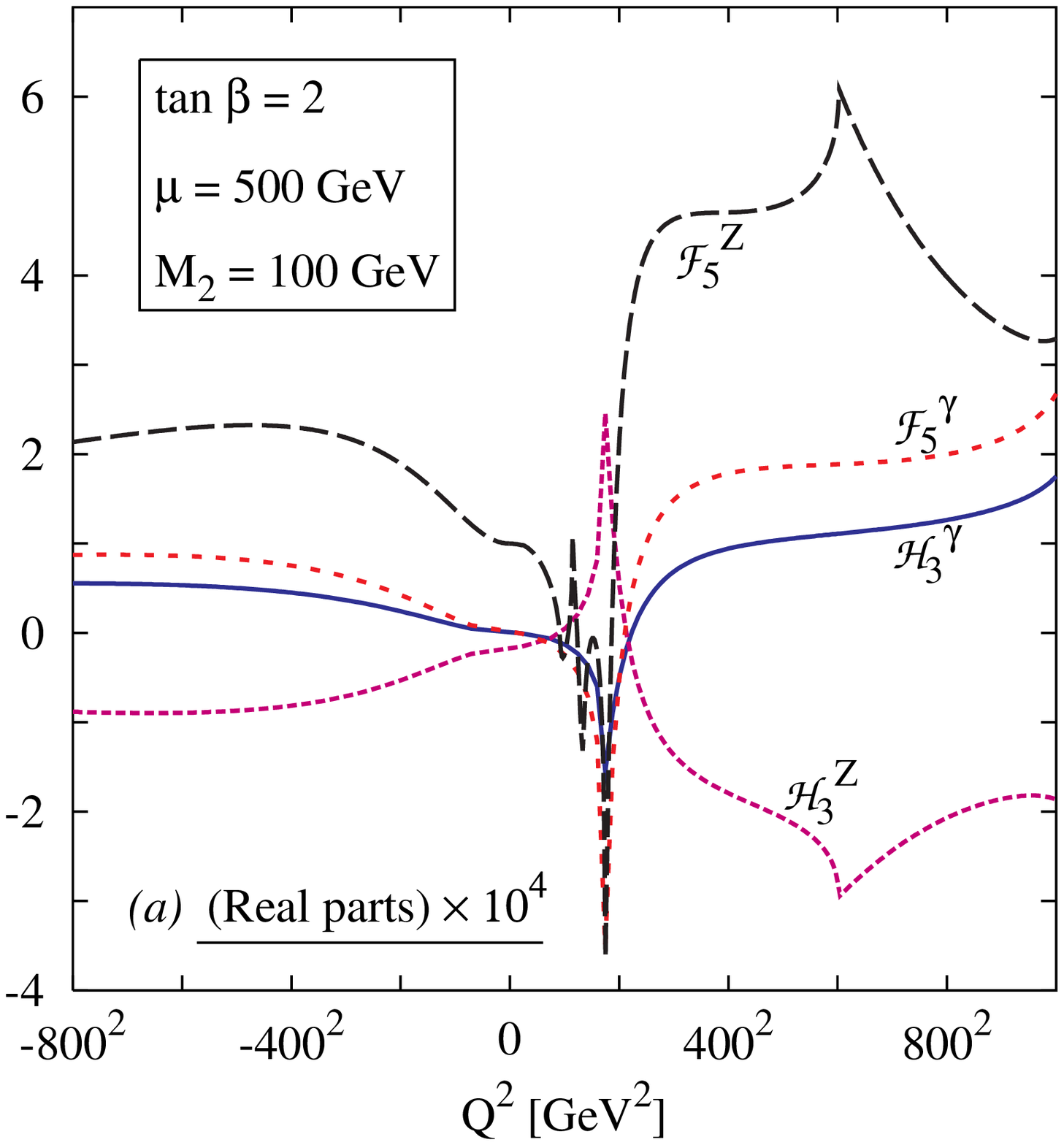} 
\vspace*{-0.0cm}
\hspace*{-0.5cm}
\epsfxsize=7.0cm\epsfysize=9.0cm
\epsfbox{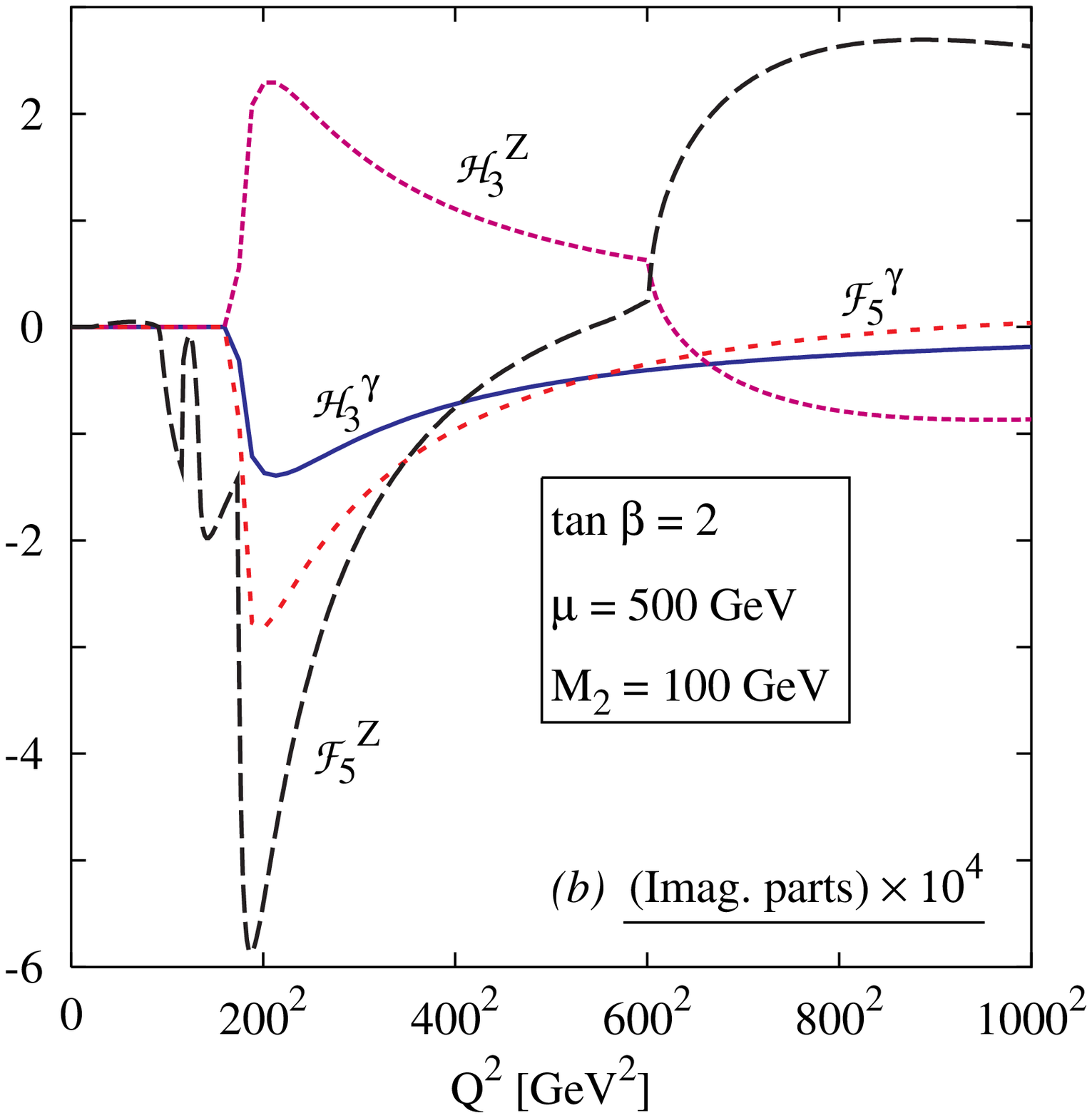}
\vspace*{-2.0cm}
}

\centerline{
\epsfxsize=7.0cm\epsfysize=9.0cm
\epsfbox{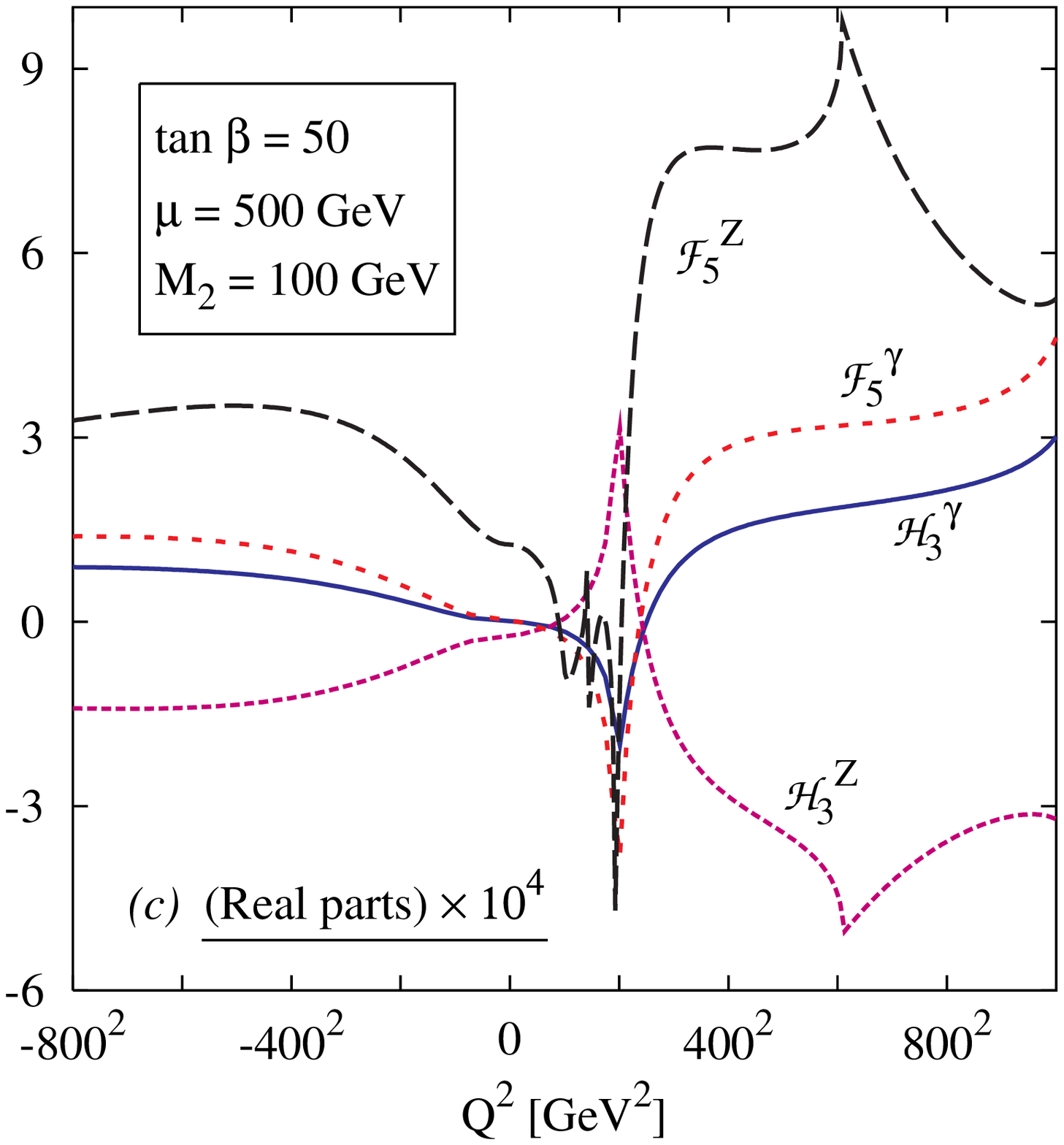} 
\vspace*{-0.0cm}
\hspace*{-0.5cm}
\epsfxsize=7.0cm\epsfysize=9.0cm
\epsfbox{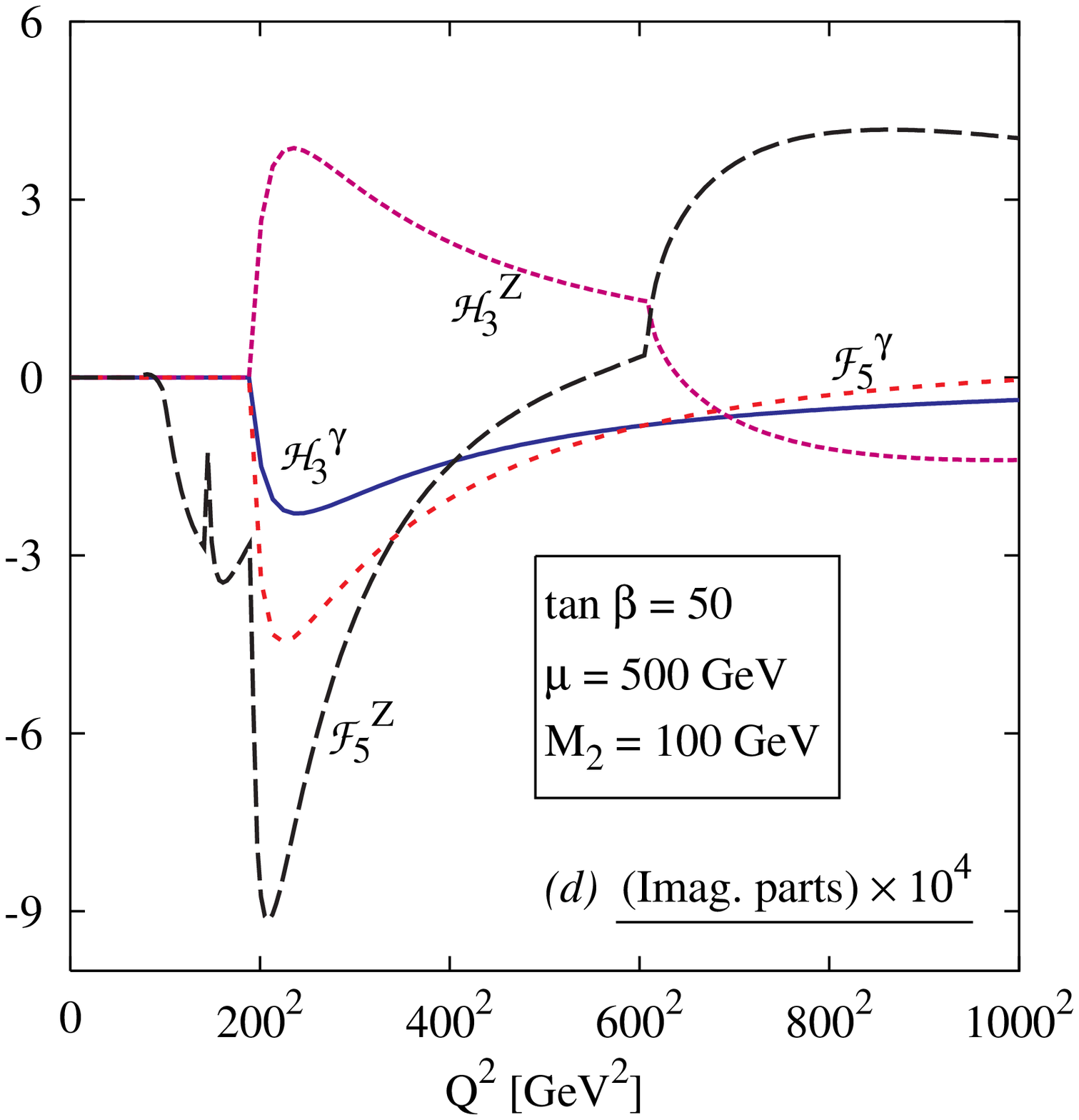}
\vspace*{-0.0cm}
}

\caption{\em The $Q^2$ dependence of the purely supersymmetric 
             contribution (within the MSSM) 
             to the four non-zero form-factors. }
      \label{fig:mssm_rts}
\end{figure}
As can be seen from Fig.~\ref{fig:mssm_rts}, the $Q^2$ behaviour is
quite analogous to that within the SM.  The size of the contribution
as well as the positions of the thresholds are of course different on
account of the quantum numbers and the masses being different.  For
example, all of the form-factors exhibit the expected\footnote{The
other threshold lies beyond the scale of the graphs.}  threshold
behaviour at $Q^2 = 4 M^2_{\tilde\chi^{+}_1}$. In addition, $\cFZ$ and
$\cZ3$ show a second threshold kink at $Q^2 =
(M_{\tilde\chi^{+}_1}+M_{\tilde\chi^{+}_2})^2$.  The other two
form-factors do not exhibit corresponding kinks due to the presence of
an off-shell photon which couples only to identical charginos.  $\cFZ$
contains effects from neutralinos as well. However the charginos
always dominate over the neutralinos except at the thresholds $2
M_{\tilde\chi^{0}_1}$ and $M_{\tilde\chi^{0}_1}+M_{\tilde\chi^{0}_2}$.
The fact, for our choice of referral parameters, of all the charginos
and neutralinos having a mass larger than $m_Z/2$ has an obvious
consequence.  Referring back to the arguments in section~\ref{sec:SM},
it is easy to see that the supersymmetric contribution to the
imaginary parts of the form-factors vanishes identically for $Q^2 <
0$.

\subsection{Dependence on the parameter space}
\begin{figure}[htb]
\vspace*{-2.5cm}
\centerline{
\epsfxsize=8.0cm\epsfysize=10.0cm
\epsfbox{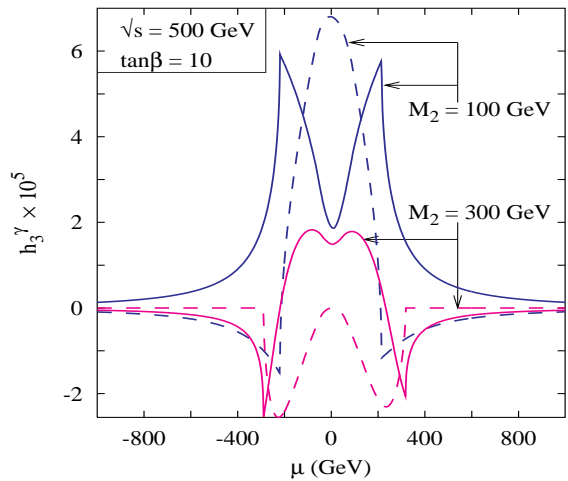} 
}
\caption{\em The dependence of the supersymmetric 
             contribution to $h_3^\gamma$ on the 
             Higgsino mass parameter $\mu$, for two particular 
             values of $M_2$.  The solid and dashed lines refer 
             to the real and imaginary parts of the form-factor.
        }
      \label{fig:mu_dep}
\vspace*{-0.2cm}
\end{figure}

To efficiently extract the parameter space dependence, it is useful to
fix the momenta and so, in this section, we shall assume $Q^2 = (500
\gev)^2$---the popular choice for a linear collider center of mass
energy---with the other two bosons being on mass-shell.  We still are
left with $\mu$, $M_2$ and $\tan \beta$. As Fig.~\ref{fig:mssm_rts}
has already shown, the dependence on the last mentioned is
quantitative rather than qualitative. Hence we shall keep it fixed at
the intermediate value $\tan \beta = 10$. In Fig.~\ref{fig:mu_dep}, we
exhibit the $\mu$ dependence of $h_3^\gamma$ for two particular values
of $M_2$.
Let us concentrate first on the imaginary part for $M_2 = 300 \gev$.
The two thresholds $\mu \approx - 290 \gev$ and $\mu \approx 318 \gev$
represent the points beyond which the lighter chargino becomes heavier
than $250 \gev$ and hence unable to contribute to the absorptive
part. Understanding the behaviour for small $|\mu|$ takes a little
more work. In this region, the lighter chargino is mainly a
higgsino. Looking at eqns.(\ref{the_combins}) and (\ref{the_O's}), it
then becomes clear that the bulk of the contribution comes from terms
proportional to the chargino mass. Consequently, a small $|\mu|$
implies a small imaginary part of the form-factor. For $M_2 = 100
\gev$, on the other hand, the gaugino state does contribute
significantly.  With the gaugino and higgsino gauge couplings being
different, the interplay between the two is very crucial. This is what
is responsible for the steep fall.

\begin{figure}[htb]
\vspace*{-2.7cm}
\centerline{
\epsfxsize=7.0cm\epsfysize=9.0cm
\epsfbox{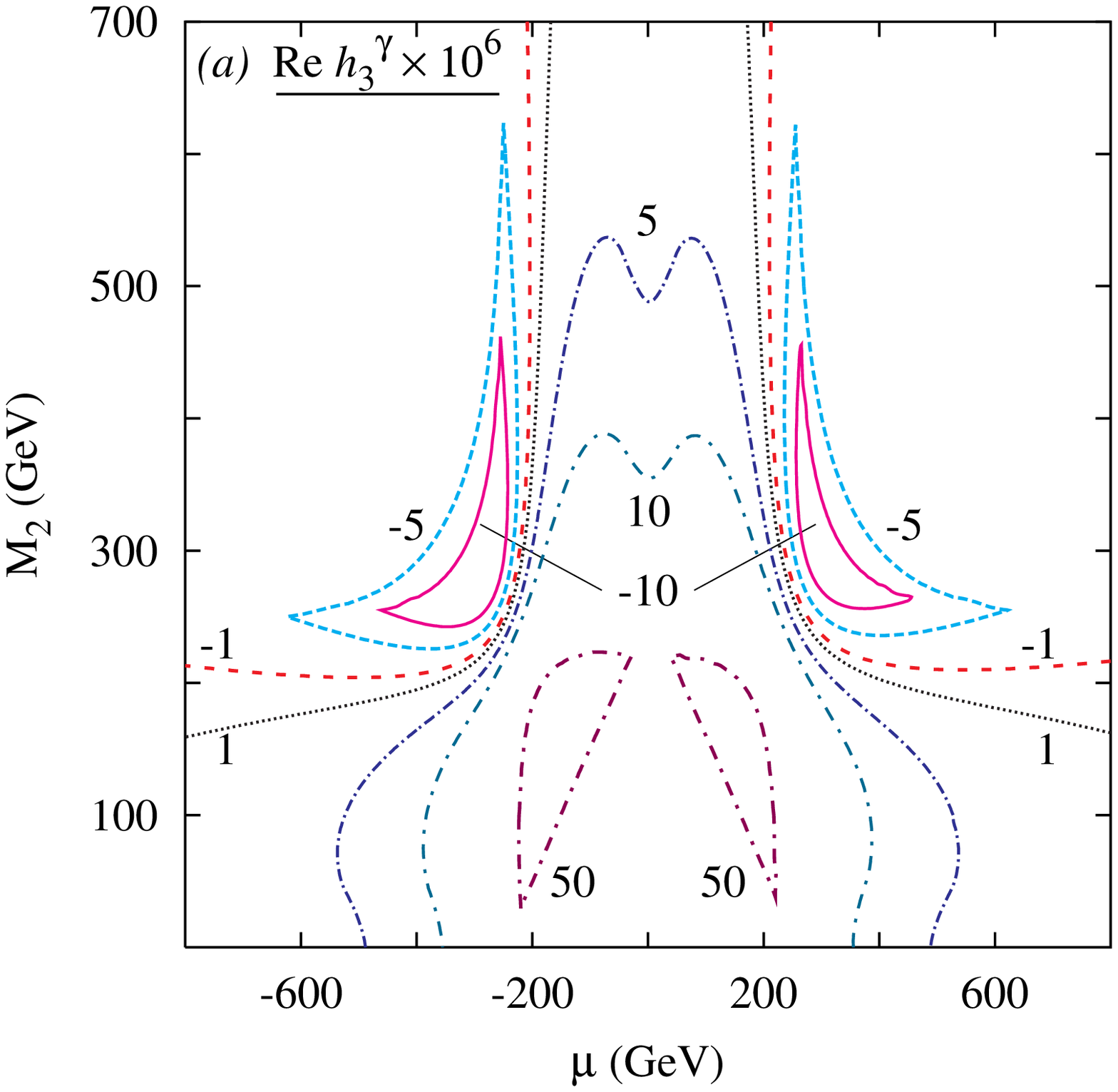} 
\vspace*{-0.0cm}
\hspace*{-0.5cm}
\epsfxsize=7.0cm\epsfysize=9.0cm
\epsfbox{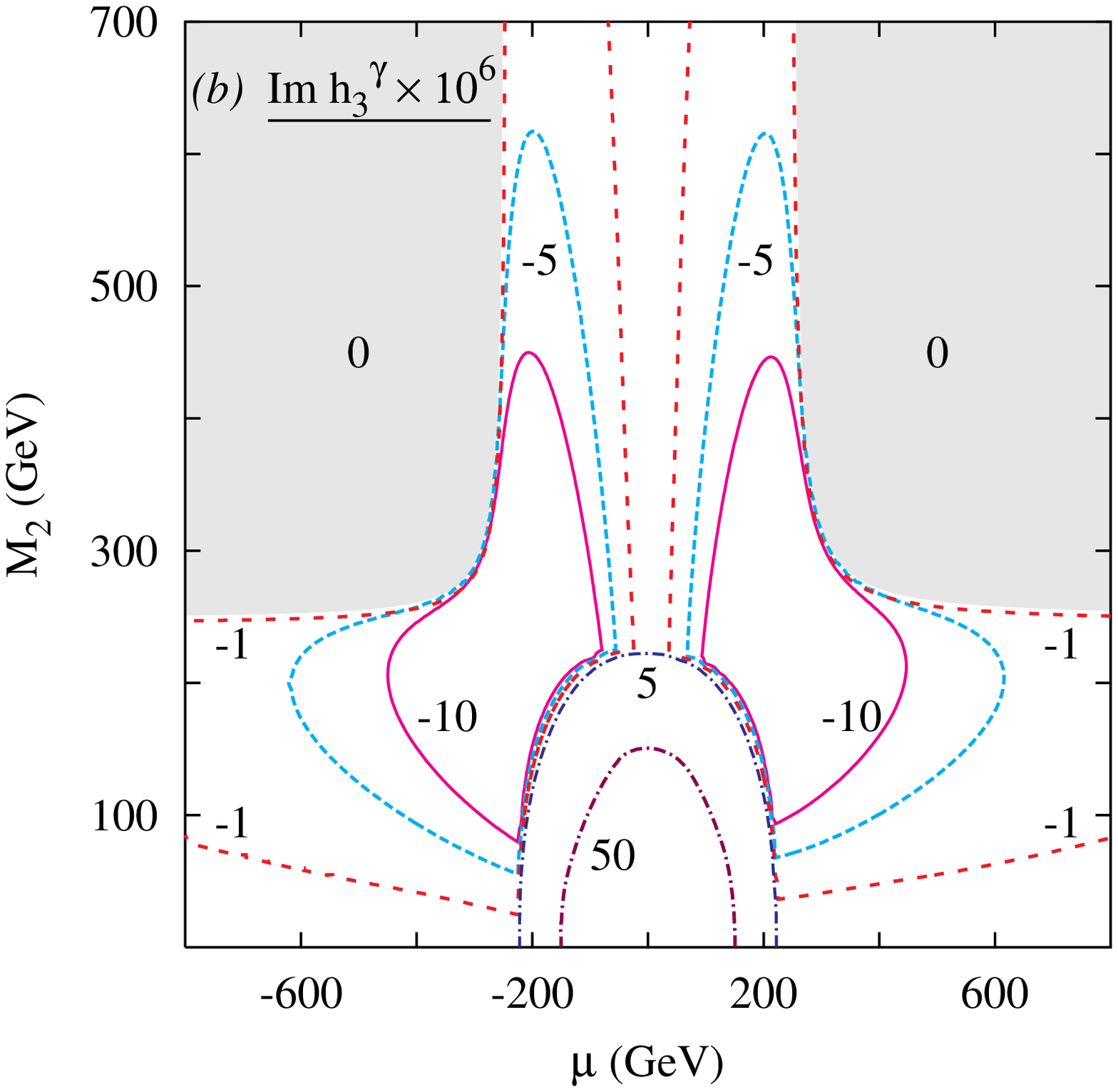}
\vspace*{-2.7cm}
}

\centerline{
\epsfxsize=7.0cm\epsfysize=9.0cm
\epsfbox{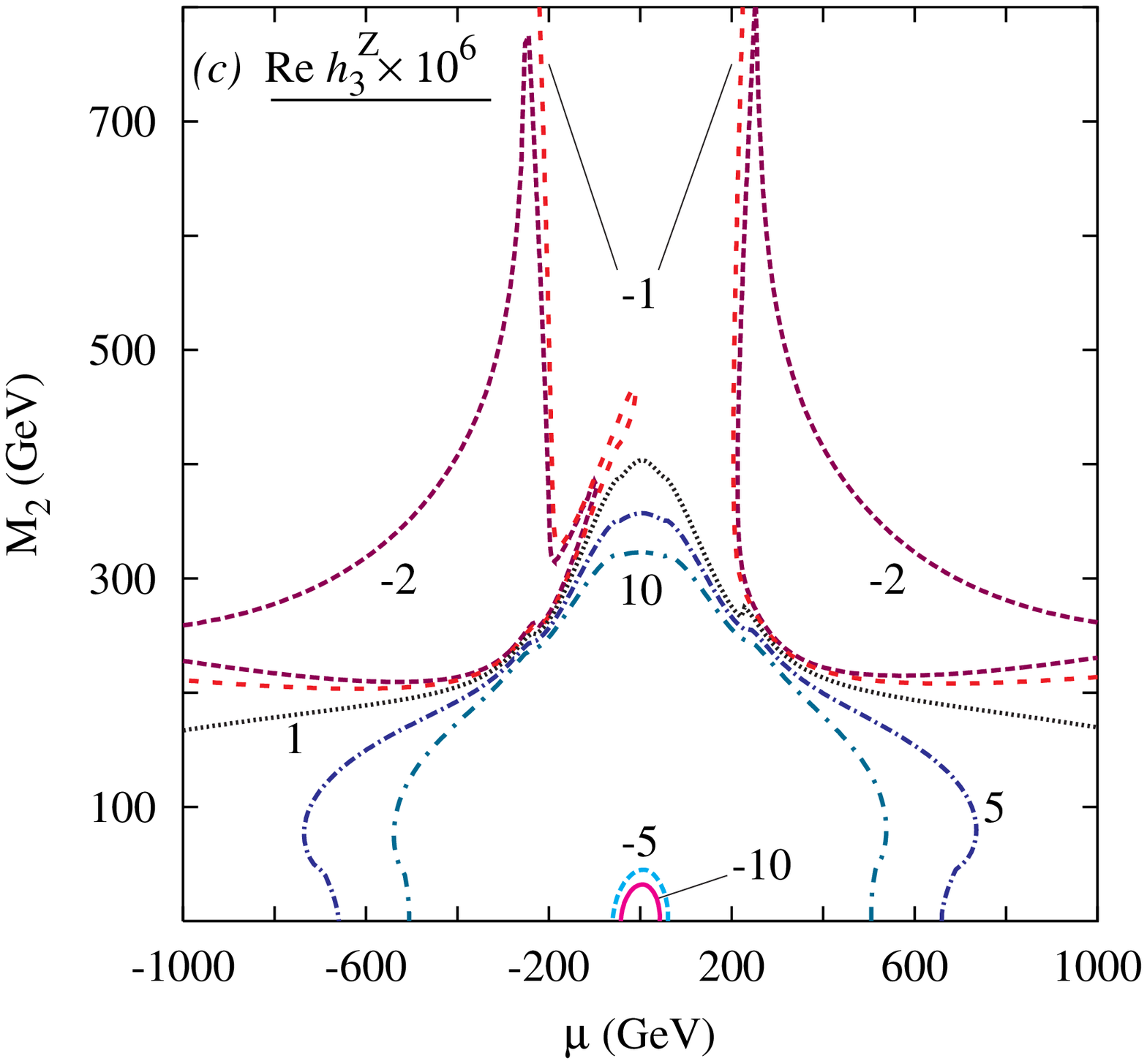} 
\vspace*{-0.0cm}
\hspace*{-0.5cm}
\epsfxsize=7.0cm\epsfysize=9.0cm
\epsfbox{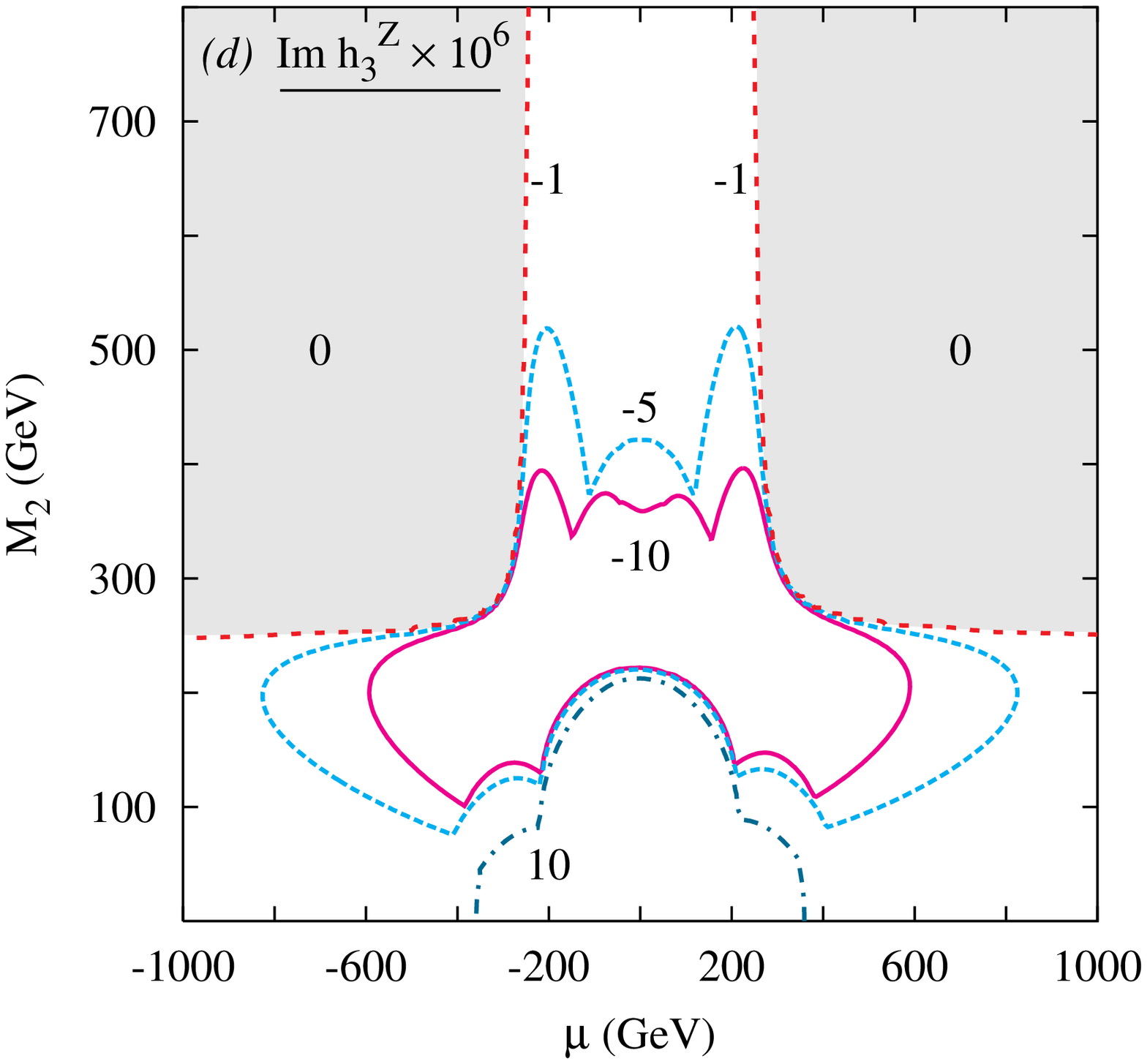}
\vspace*{-0.3cm}
}
\vskip 10pt
\caption{\em Contours for constant real and imaginary  
             parts of the supersymmetric contribution to the 
             form-factors {\em (}$\tan \beta = 10${\em )}. In graphs 
             {\em (b)} and {\em (d)}, the shaded area denotes 
             the parameter space where the imaginary parts 
             of the form-factors are identically zero.
        }
      \label{fig:mssm_cont}
\end{figure}

The same steep slope also implies that contour levels (for the
imaginary part) in the $(\mu, M_2)$ plane would be rapidly
changing. This is reflected in the plots of Fig.~\ref{fig:mssm_cont}, 
where we exhibit the behaviour of $h_3^\gamma$ and $h_3^Z$ 
as we vary either or both of $\mu$ and $M_2$ 
while maintaining  $\tan\beta=10$. 

Let us consider the $\gamma\gamma Z$ vertex where at a time only one
species of the charginos can flow inside the loop. Regarding the
parameter space dependence, if one increases $M_2$ keeping $\mu$ and
$\tan\beta$ fixed, then the heavier chargino ($\tilde\chi^{+}_1$)
becomes more gaugino-like and correspondingly, the lighter one
($\tilde\chi^{+}_2$) becomes more higgsino-like. So as $M_2$
increases, the $Z\tilde\chi^{+}_1\tilde\chi^{-}_1$ coupling becomes
stronger (remember the $Z$ couples to two $W^\pm$'s in the SM) and the
$Z\tilde\chi^{+}_2\tilde\chi^{-}_2$ coupling becomes weaker. However,
the diagram with $\tilde\chi^{+}_1$ flowing inside the loop suffers
from large propagator suppressions in comparison with the other one
containing $\tilde\chi^{+}_2$. Ultimately the magnitude of the total
MSSM loop contribution to $h_3^\gamma$ decreases. If we choose to
increase $\mu$ fixing $M_2$ and $\tan\beta$, $\tilde\chi^{+}_1$
becomes more higgsino-like. Here the diagram containing
$\tilde\chi^{+}_2$ wins from both the strength of
$Z\tilde\chi^{+}_2\tilde\chi^{-}_2$ coupling and the propagator
suppressions.  So in this case the total one-loop contribution falls
off as well.

The parameter space dependence of other form-factors are more
complicated as in those cases different types of charginos can
co-exist inside the loop. However the basic reasoning is the same as
above.  As said earlier, the neutralinos contribute only to
$f_5^Z$. The gaugino-like neutralinos can not contribute as there are
no trilinear neutral gauge boson vertices at the tree level. Only the
higgsino-like neutralinos contribute. However, as as the 
chargino contributions dominate over those from the neutralinos
for most of the parameter space, we desist from discussing this
issue any further.

\section{Off-shell TNGBVs}

Until now, we have dealt with the case wherein only one out of the 
three gauge bosons is off-shell. When more than one of them are
off-shell, the form-factors are modified in two different ways. As the
analysis of section~\ref{sec:generic} indicates, and as we 
shall shortly show, additional form-factors
are possible. Apart from this, even the ones that we
have considered are modified in a significant way. We examine 
the latter consequence first. 

\subsection{Modifications to ${\cal H}^{\gamma/Z}_3$ and ${\cal
  F}^{\gamma/Z}_5$}

It is clear that now we can no longer talk in terms of $h^{\gamma/Z}_3$ {\em
  etc}. To take into account the explicit dependence on $p_1^2$ and $p_3^2$ as
in eqns.(\ref{g_gst_Z_b},\ref{H3Z_b},\ref{f5g_b} and \ref{f5Z_b}), it is 
imperative that we consider the full form-factors 
${\cal H}^{\gamma/Z}_3$ and ${\cal  F}^{\gamma/Z}_5$. 
Of course, over and above this explicit dependence on
$p_{1,3}^2$, an additional dependence appears through the arguments of the
Passarino-Veltman functions.

\begin{figure}[htb]
\centerline{
\epsfxsize=7.0cm\epsfysize=6.0cm
\epsfbox{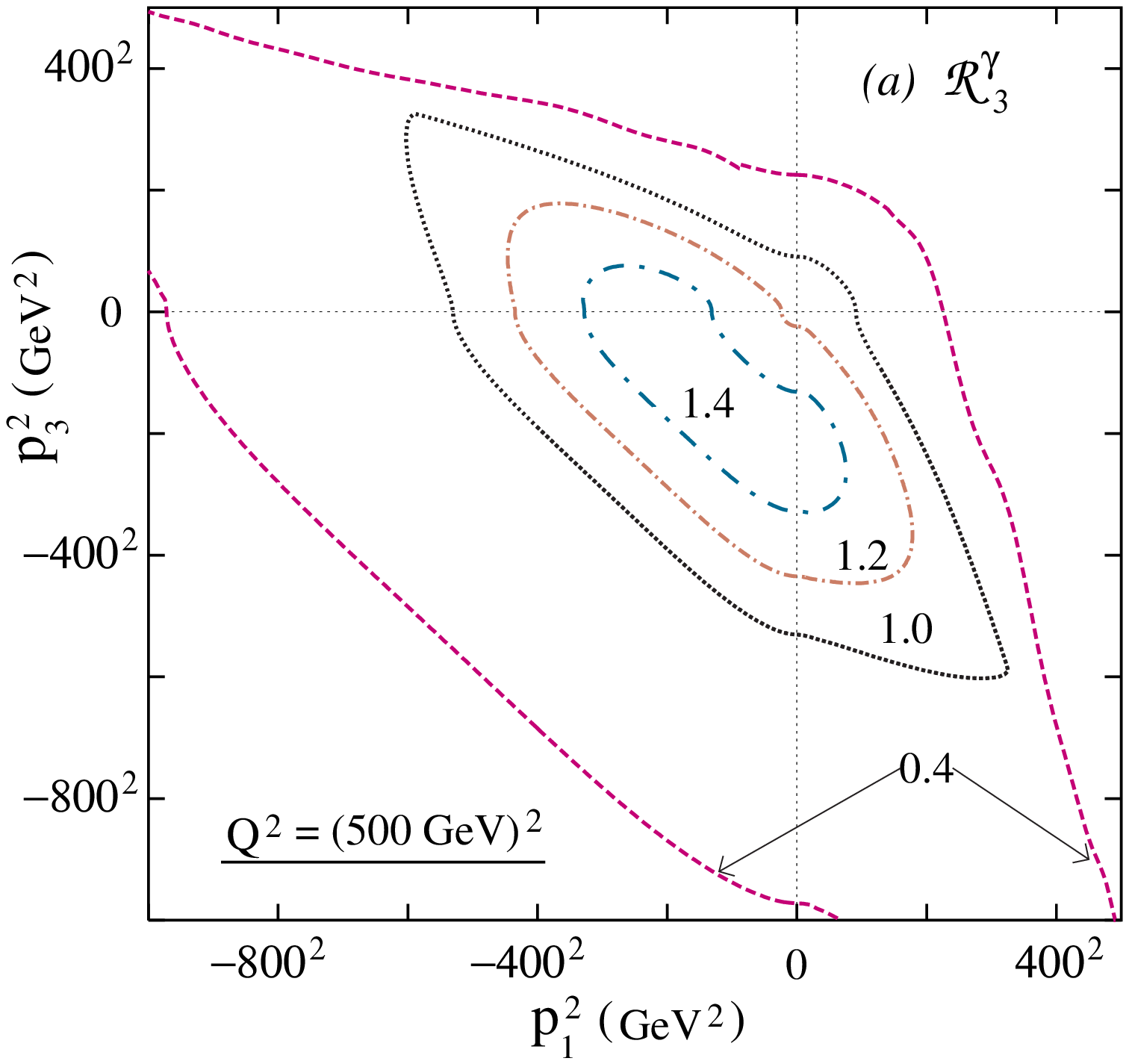} 
\vspace*{-0.0cm}
\hspace*{-0.5cm}
\epsfxsize=7.0cm\epsfysize=6.0cm
\epsfbox{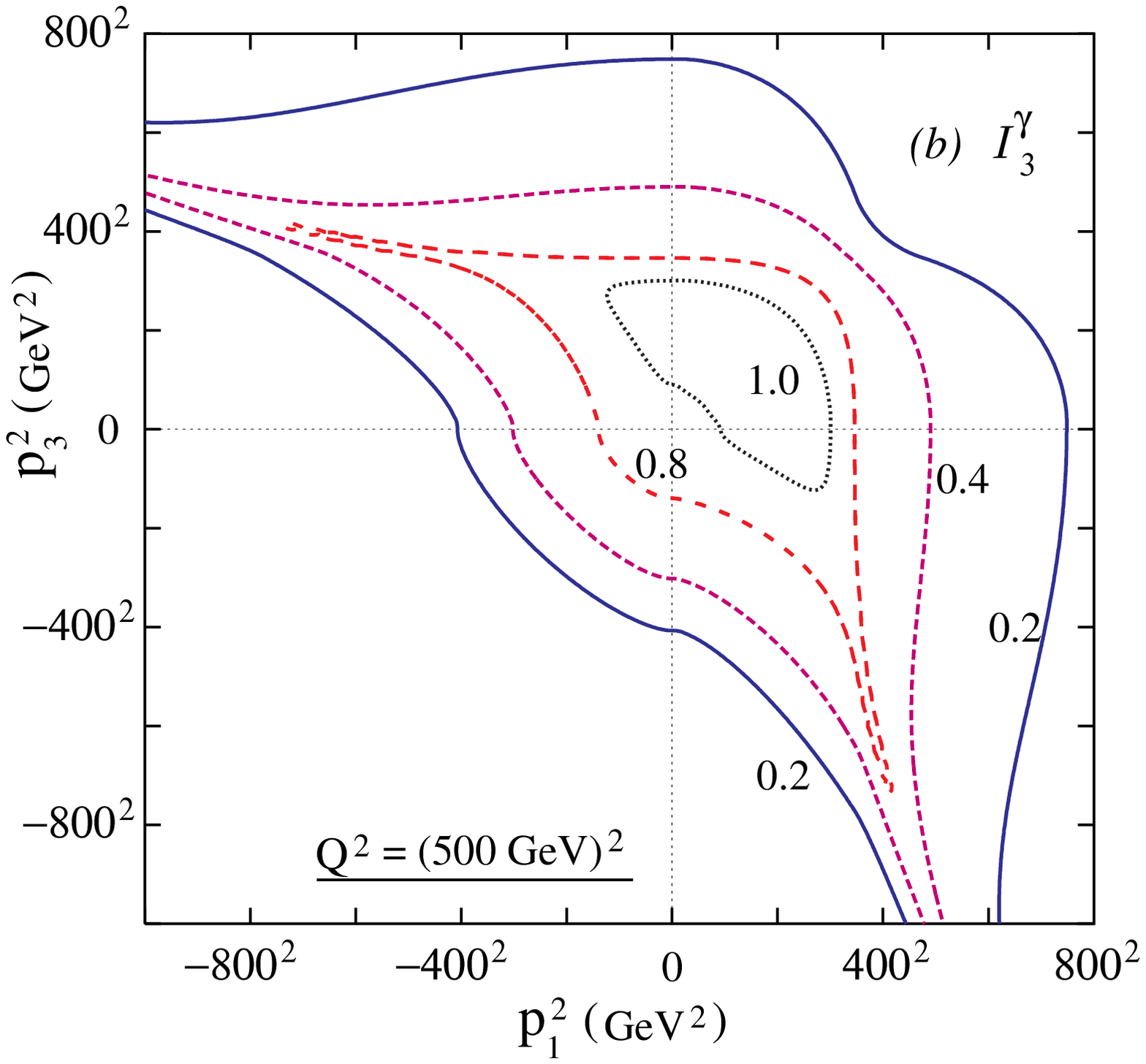}
\vspace*{-3.6cm}
}
\vspace*{0.3cm}\hspace*{-2.0cm}
\centerline{
\epsfxsize=6.8cm\epsfysize=10.0cm
\epsfbox{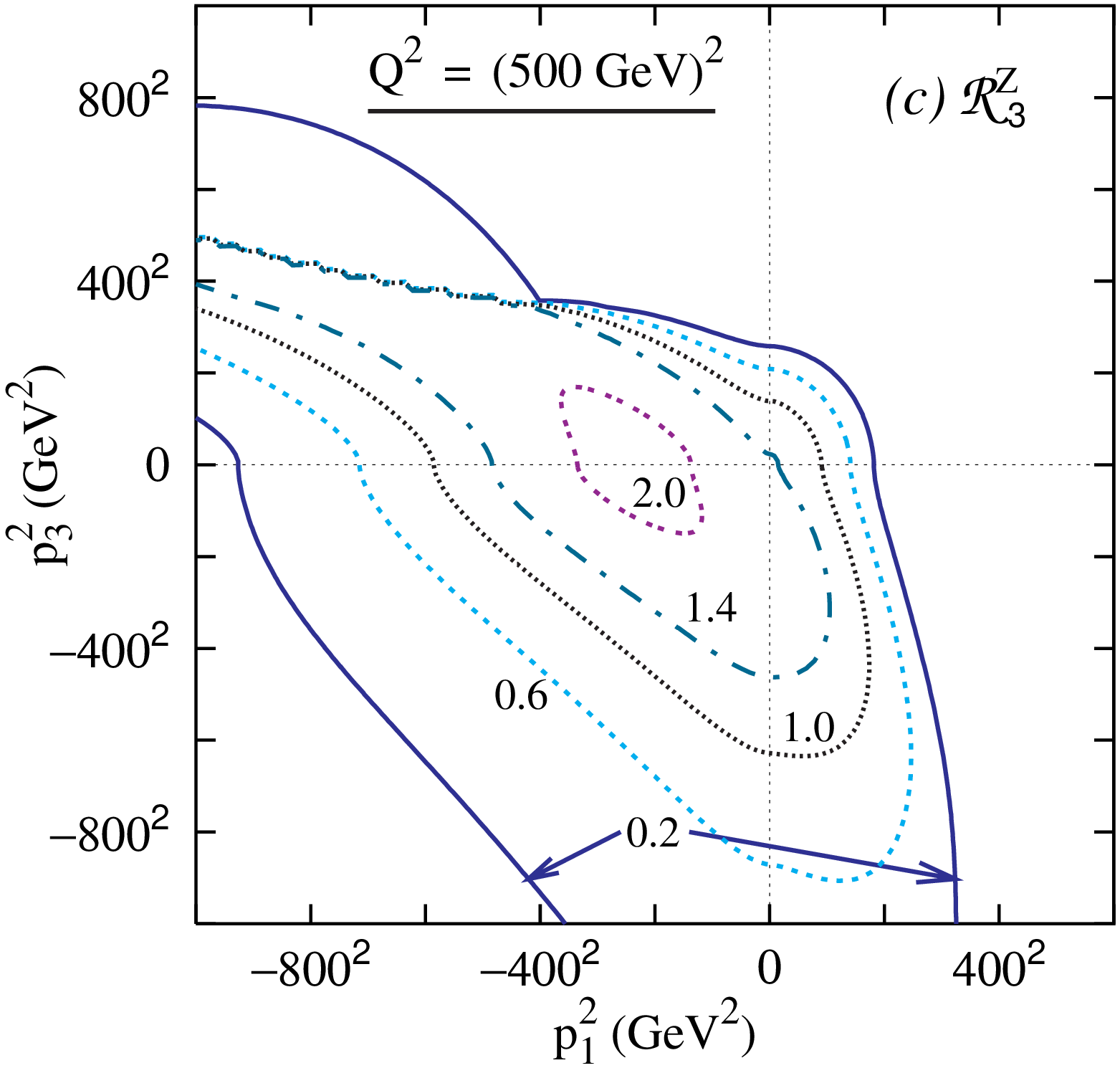} 
\vspace*{0.0cm}
\hspace*{-0.3cm}
\epsfxsize=6.8cm\epsfysize=10.0cm
\epsfbox{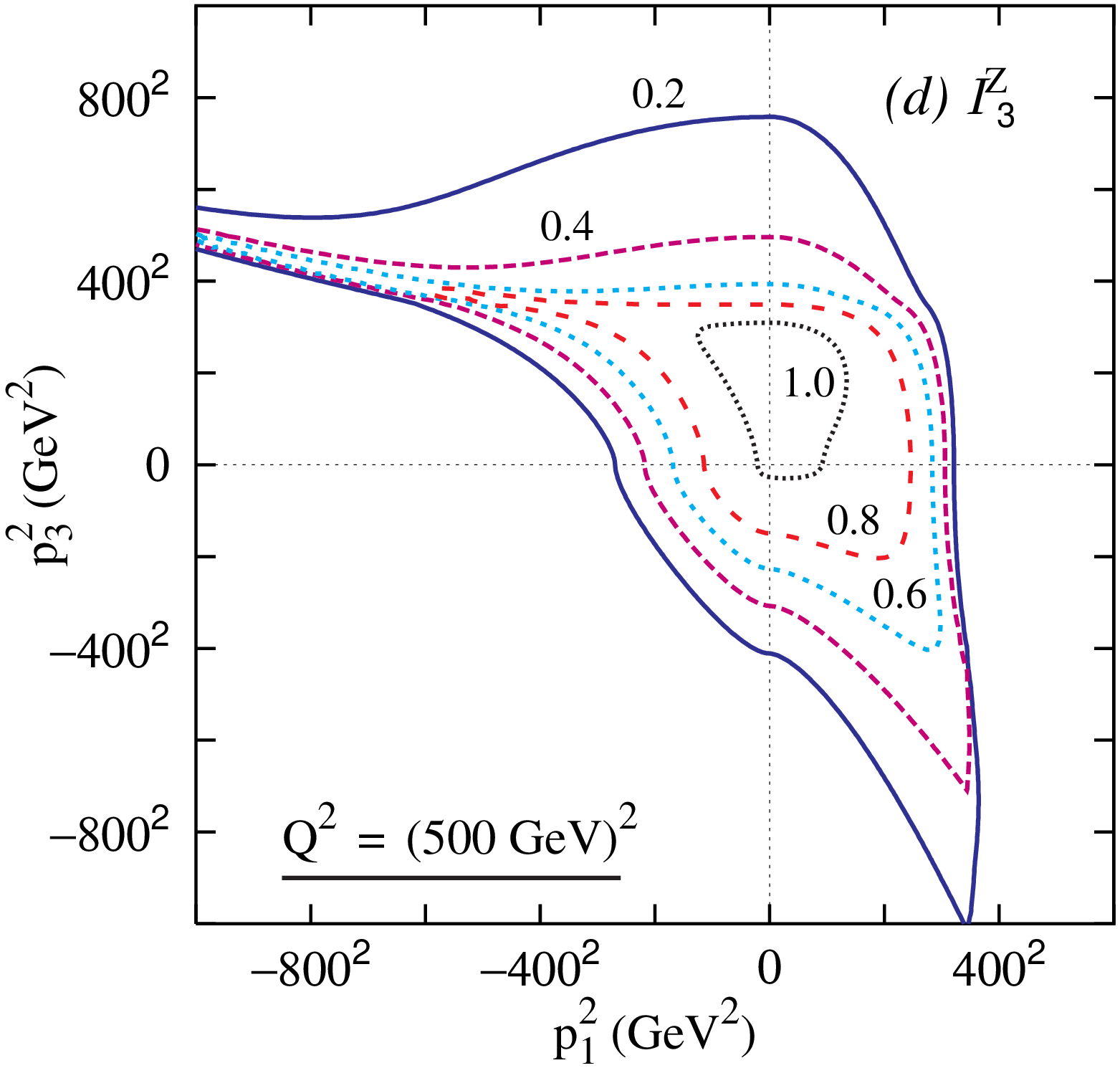}
\vspace*{-0.3cm}
}

\caption{\em Constant ${\cal R}$ and ${\cal I}$ contours 
             (see eqn.\ref{ofsh_defn}) within the SM 
             in the $(p_1^2, p_3^2)$ plane.
        }
      \label{fig:offshell}
\end{figure}

To quantify the changes wrought by the hitherto on-shell particles
going off-shell, we define the ratios
\be 
\barr{rcl}
{\cal R}_3^\gamma & \equiv & \dis \frac{[Re(\cg3)]_{\rm off-shell}}
                                     {[Re(\cg3)]_{\rm on-shell}} 
        \\[3ex]
{\cal I}_3^\gamma & \equiv & \dis \frac{[Im(\cg3)]_{\rm off-shell}}
                                     {[Im(\cg3)]_{\rm on-shell}} 
\earr
        \label{ofsh_defn}
\ee
where the subscripts ``on-shell'' and ``off-shell'' are self-explanatory.
Analogous definitions hold for ${\cal R}_3^Z$ and ${\cal I}_3^Z$.  In
Fig.~\ref{fig:offshell} we present the contours for constant ${\cal
  R}_3^{\gamma/Z}$ and ${\cal I}_3^{\gamma/Z}$ in the $(p_1^2, p_3^2)$ plane
for a fixed value $p_2^2 = (500 \gev)^2$. 

\begin{figure}[htb]
\vspace*{-4cm}
\centerline{
\epsfxsize=8.0cm\epsfysize=10.0cm
\epsfbox{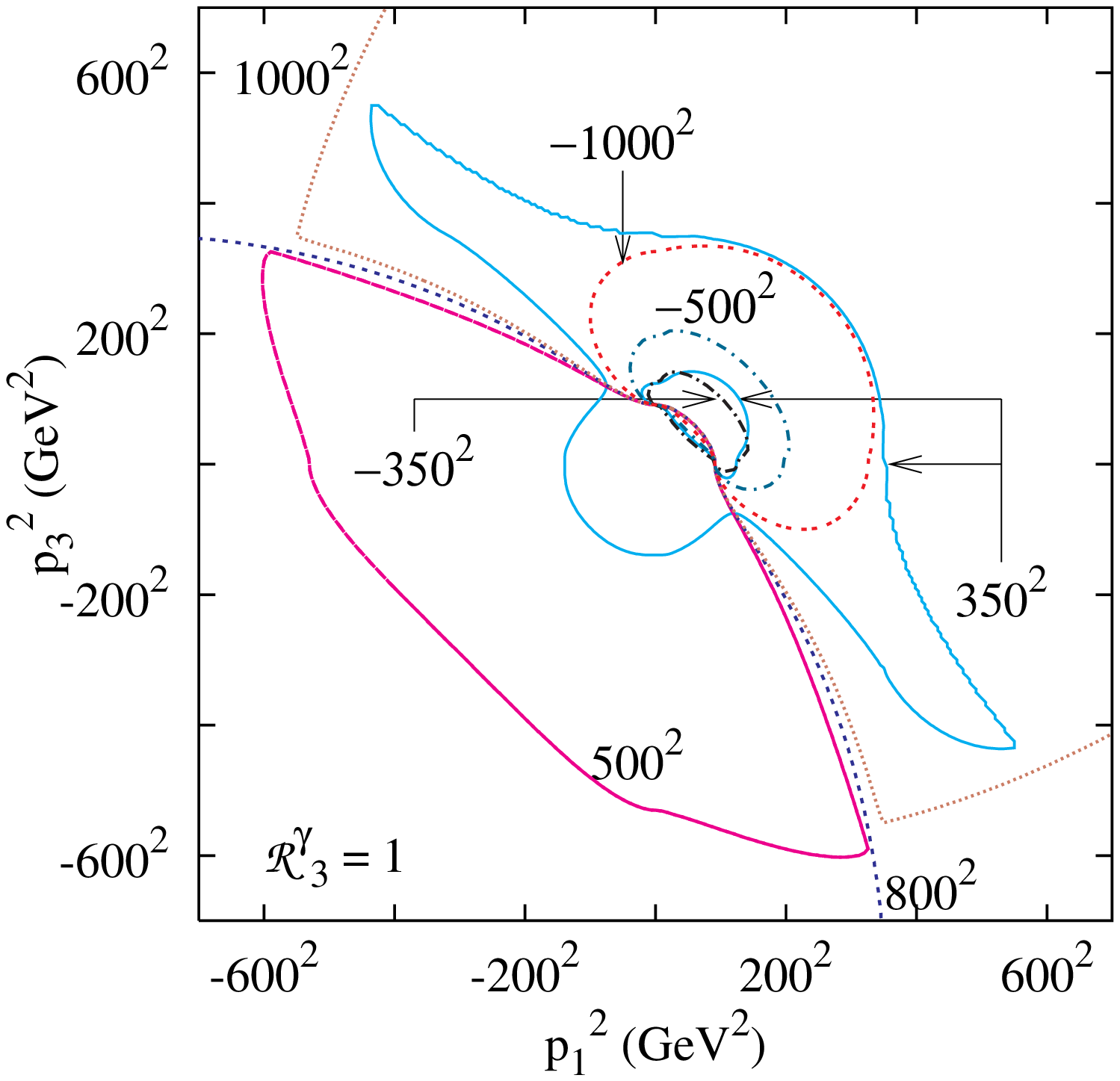} 
\vspace*{-0.0cm}
\hspace*{-0.5cm}
\epsfxsize=8.0cm\epsfysize=10.0cm
\epsfbox{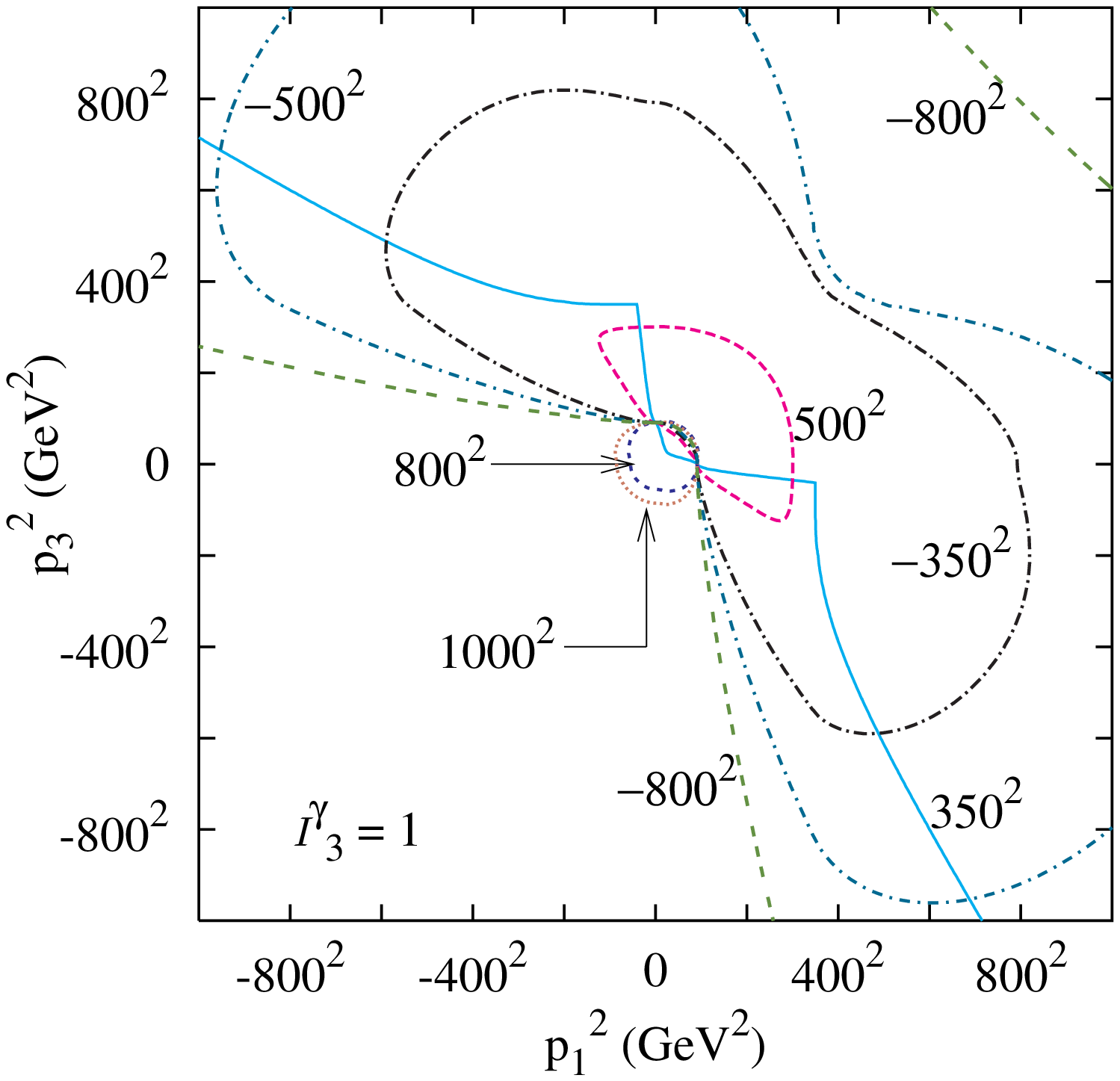}
\vspace*{-0.3cm}
}
\caption{\em {\em (a)} Contours for ${\cal R}_3^\gamma = 1$ 
                in the $(p_1^2, p_3^2)$ plane for different values of 
                $p_2^2$ (shown as legends). {\em (b)} Similar contours
                for ${\cal I}_3^\gamma = 1$. All calculations are 
                within the SM.
        }
      \label{fig:offsh_qsq}
\end{figure}

As for the on-shell form-factors, it is interesting to consider the dependence
on $p_2^2$. Naively, one would expect a behaviour broadly similar to that in
Fig.~\ref{fig:sm}. That this is indeed so can be seen by an examination of
Fig.~\ref{fig:offsh_qsq}.  At a first glance though, this assertion might seem
unfounded since these figures do not seem to show all the features of
Fig.~\ref{fig:sm}, notably the kink. However, if one were to draw more
contours for intermediate values, at the cost of cluttering the graph, such
features spring out immediately.

\subsection{``Truly off-shell'' form-factors}

To identify all the possible additional form-factors that may arise 
when more than one of the bosons is off mass-shell, 
we revert back to eqn.(\ref{master}) and restate it in a
much more compact and simplified form. 
Gauge invariance implies that any term proportional
to a photon momentum can be dropped without loss of generality. While
this argument is not applicable for the $Z$, a very similar one
holds. As long as the $Z$ couples only to light fermions ($m_f \ll
\sqrt{s}$)---as is the case almost always---current conservation
implies that terms proportional to the $Z$ momentum may be dropped as
well. Within this approximation, then, either of $B_2$ and $\bar B_2$
are irrelevant.  Restricting our analysis to one loop order further
constrains ${\bar B_3}$ to be equal to $B_3$, vide
eqn.(\ref{eq:loop}). Therefore, to this order,
\[
\cB3 p_{1 \mu} + \bar \cB3 p_{2 \mu}= \cB3 (p_1 + p_2)_\mu 
            = - \cB3 p_{3 \mu}
\]
Once again, this term may be dropped by virtue of 
current conservation. Thus, to one-loop, the generic 
TNGBV may be parametrized as 
\be
 \Gamma_{\alpha \beta \mu} (p_1, p_2 ; p_3) 
= \epsilon_{\alpha \beta \mu \eta} (B_1  p_1^\eta - \bar B_1  p_2^\eta)  
        \label{mast} 
\ee
and, hence, there are at best two form-factors for each combination 
of the vector bosons. Let us now consider each in turn.
\begin{itemize}
\item \underline{$\gamma \gamma Z$ vertex}

        Of the two form-factors in eqn.~(\ref{mast}), 
the first, namely $\cB1$,  is simply related to ${\cal H}_3^\gamma$ 
as defined for the on-shell case (see eqn. 3b). 
However, when the other gauge bosons are off-shell too, 
we have an additional form-factor, namely, 
\be 
        {\cal H}_5^\gamma = i \bar{\cB1}.  
\ee
While the variation of ${\cal H}_5^\gamma$ with each of the external 
momenta could be of interest, we restrict ourselves to the special 
situation wherein a comparison with ${\cal H}_3^\gamma$ is the easiest,
namely $p_2^2 = (500\gev)^2$. This particular value is of interest as it 
is widely considered to be the choice of center of mass energy for the 
first generation linear collider. 
In Fig.~\ref{H5g} we plot the variation of 
${\cal H}_5^\gamma$ with $p_1^2$ for different 
values of $p_3^2$. As earlier, it is more instructive to concentrate 
on the imaginary part of the form-factor since the real part 
can then be obtained using a dispersion relation. 
Expectedly,  in the imaginary part, we see a kink at
$p_1^2=(2\,m_t)^2$, with a corresponding dip in the real 
parts\footnote{Similar kinks (dips) occur for other fermion 
        thresholds, but these are too small and close to the origin 
        to be visible in the scale of the graph.}.
Note that unlike in the case of Section \ref{sec:SM}, for a 
given fermion, the 
diagram may now be cut in more than one way and 
an imaginary contribution obtained. 
In Fig.~\ref{H5g}, this feature manifests itself rather dramatically 
for $n=5$. Concentrating on the top-mediated diagram, a 
simple calculation shows that for $p_1^2 \simeq - (900 \gev)^2$,
each of the three top-quarks could be on-shell. The 
kink thus corresponds to one particular cut ceding dominance to 
another.
%
\begin{figure}[htb]
\vspace*{-6.0cm}
\hspace*{-1.1cm}
\centerline{
\epsfxsize=9.5cm\epsfysize=13.0cm
\epsfbox{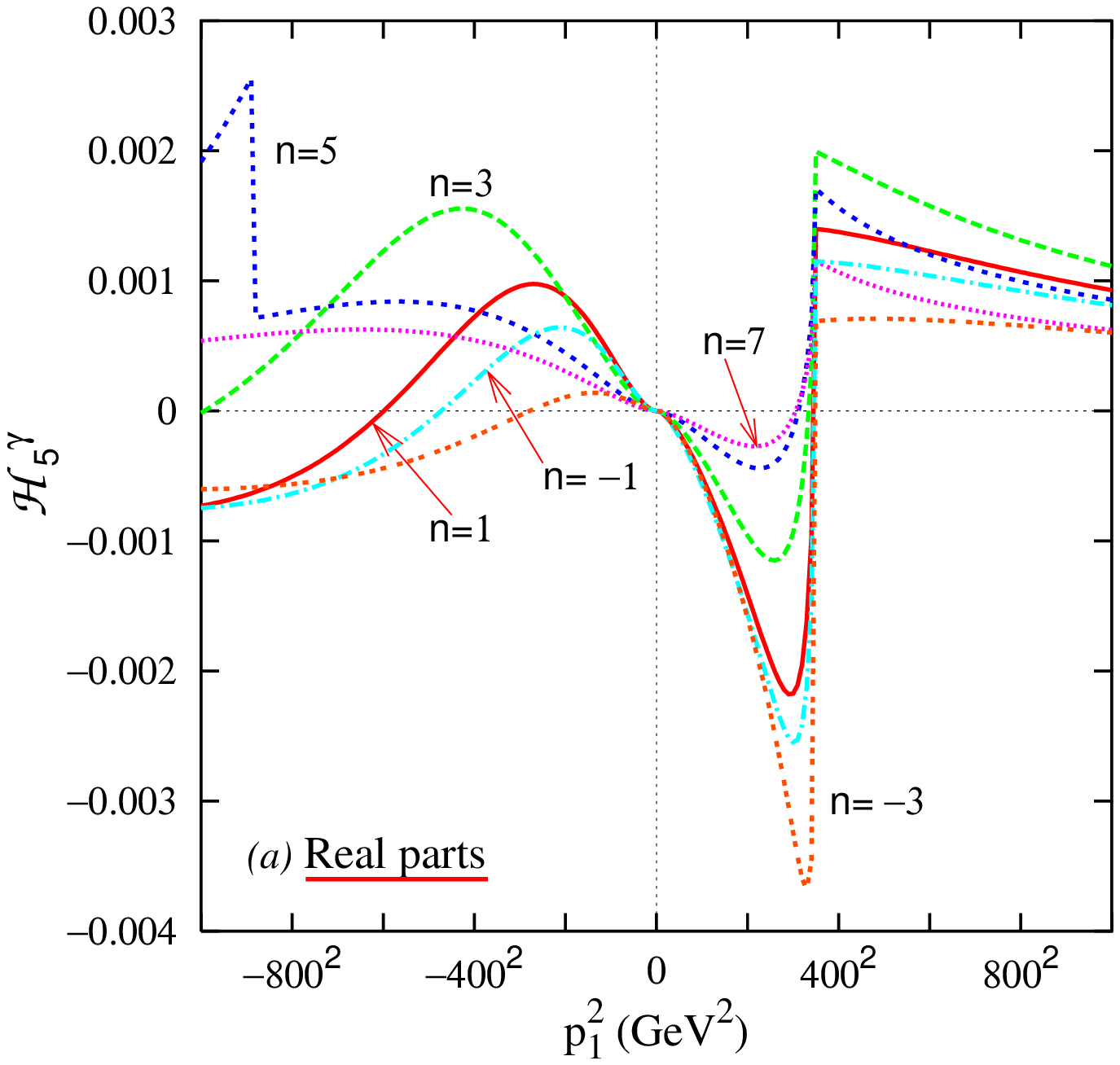} 
\vspace*{-0.0cm}
\hspace*{-1.5cm}
\epsfxsize=9.5cm\epsfysize=13.0cm
\epsfbox{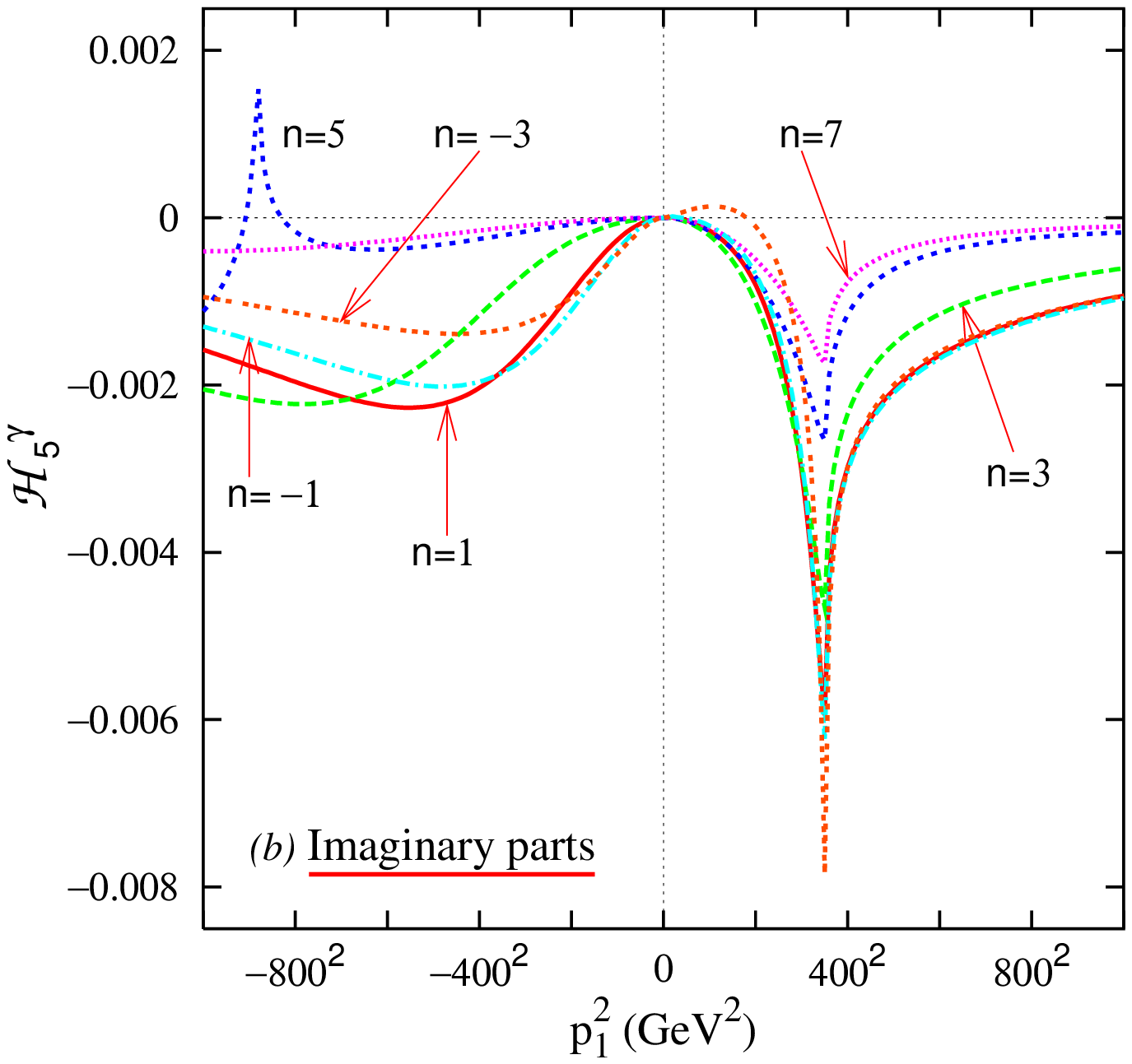}
\vspace*{-0.3cm}
}
\caption{\em The $Q^2$ dependence of the new non-zero
  form-factor ${\cal H}_5^\gamma$ within the SM for the off-shell
  $\gamma\gamma Z$ vertex for $p_2^2 = (500\gev)^2$. $p_3^2 = (n \;M_Z)^2$ for
  positive n and $p_3^2 = -\,(n \: M_Z)^2$ when n is negative.}
      \label{H5g}
\end{figure}
%

\item \underline{$Z Z \gamma$ vertex}

Before manipulating eqn.(\ref{mast}), we reexpress it as
\be
 \Gamma^{ZZ\gamma}_{\alpha \beta \mu} (p_1, p_2 ; p_3) 
    = \epsilon_{\alpha \beta \mu \eta}  {1 \over 2}\, 
             \left [   (\cB1 + \bar \cB1)  (p_1 - p_2)^\eta 
                 + ( \cB1 -  \bar \cB1 ) ( p_1 + p_2)^\eta
              \right]
\ee
On imposing gauge invariance (eqn.\ref{ZZg_gaugeinv}), this reduces 
to
\[
\Gamma^{ZZ\gamma}_{\alpha \beta \mu} (p_1, p_2 ; p_3) 
 =    \epsilon_{\alpha \beta \mu \eta}  {1\over  2}\,
  \left[\cB3  p_3^2 (p_1 - p_2)^\eta - (\cB1 -  \bar\cB1) p_3^\eta \right]
\]
These two form-factors have already been 
identified as ${\cal F}_5^\gamma$ and ${\cal H}_3^Z$ 
respectively. Thus, as expected, no additional form-factors appear even 
when we generalise to the case where all the gauge bosons are
off-shell.

\item \underline{$Z Z Z$ vertex}

We could, once again, reexpress eqn.(\ref{mast}) as in the case 
for the $Z Z \gamma$ vertex. Note that while $(\cB1 + \bar\cB1)$  corresponds
to ${\cal F}_5^Z$, the orthogonal combination is a new form-factor in its own right. Thus, we may define ${\cal F}_6^Z$, {\it via} 
\be 
{\cal F}_6^Z = i \frac{(\cB1 - \bar \cB1)}{2}  
\ee 
Obviously, ${\cal F}_6^Z$ vanishes identically for $p_1^2$ = $p_3^2$. 
And, given the symmetry of the problem, the iso-${\cal F}_6^Z$ contours 
in the $(p_1^2, p_3^2)$ plane are symmetric about this line. 
In Fig.~\ref{F6z}, we display the momentum dependence of 
${\cal F}_6^Z$ in a fashion similar to that for ${\cal H}_5^\gamma$. 
The behaviour of the curves can be understood in an analogous manner.
%
\begin{figure}[htb]
\vspace*{-6.0cm}
\hspace*{-1.1cm}
\centerline{
\epsfxsize=9.5cm\epsfysize=13.0cm
\epsfbox{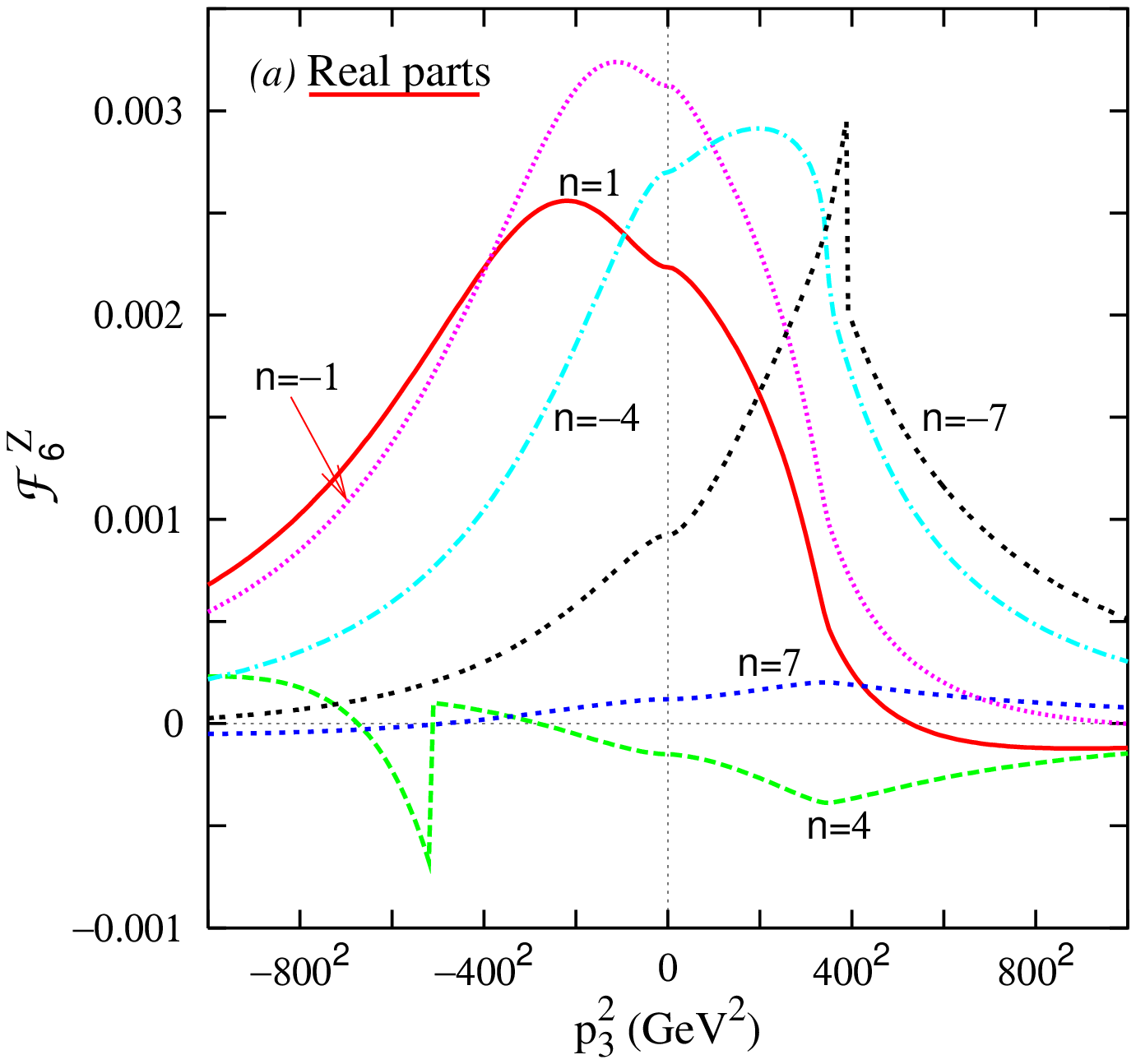} 
\vspace*{-0.0cm}
\hspace*{-1.5cm}
\epsfxsize=9.5cm\epsfysize=13.0cm
\epsfbox{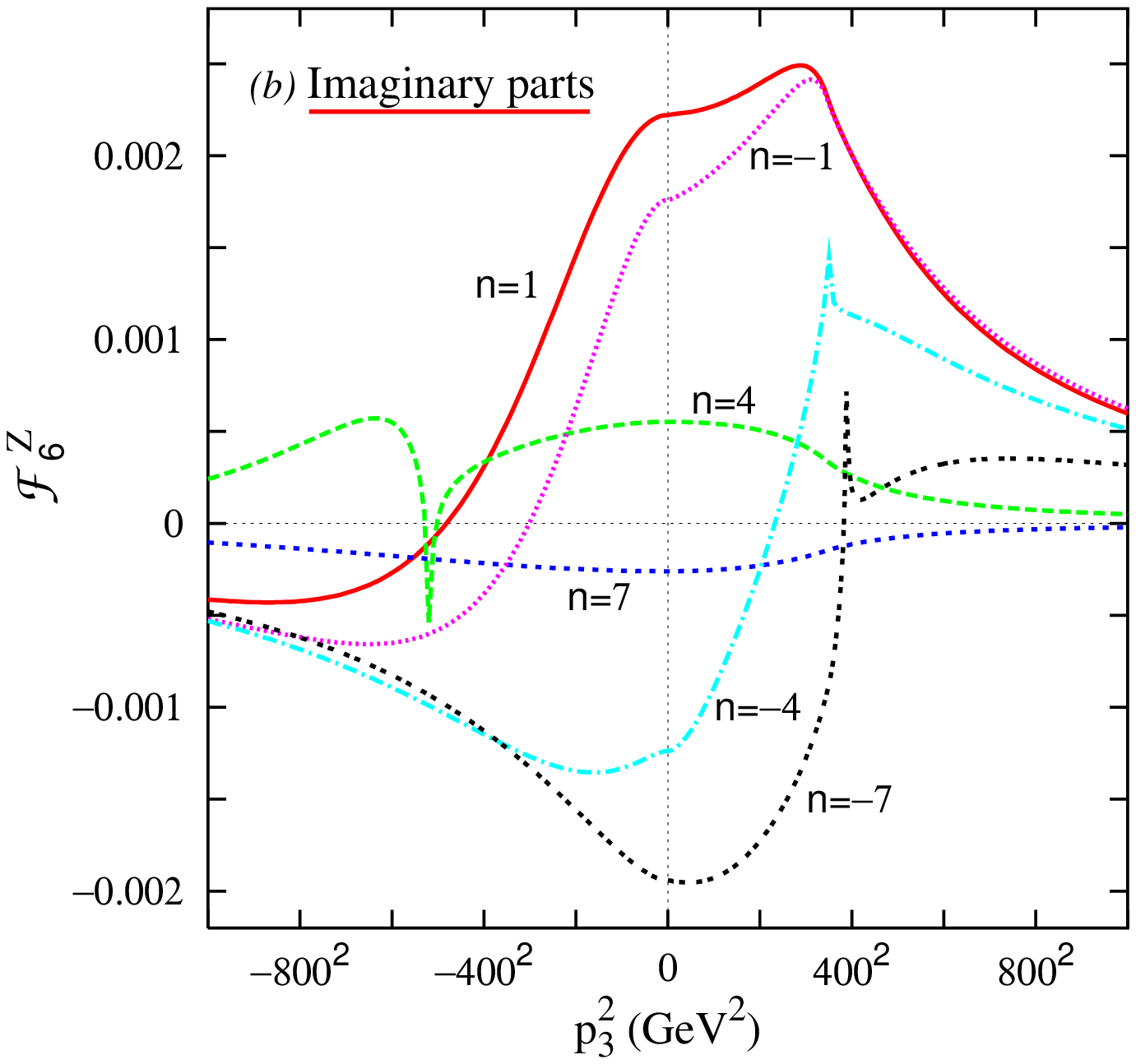}
\vspace*{-0.3cm}
}
\caption{\em The $Q^2$ dependence of the new non-zero
  form-factor ${\cal F}_6^Z$ within the SM for the off-shell $ZZZ$ vertex for
  $p_2^2 = (500\gev)^2$. $p_1^2 = (n \;M_Z)^2$ for positive n and $p_1^2 =
  -\,(n \: M_Z)^2$ when n is negative.}
      \label{F6z}
\end{figure}
%
\end{itemize}

\section{Conclusions}
In this paper we have considered the SM and MSSM contributions to
trilinear neutral gauge boson vertices namely, $\gamma\gamma Z$, $Z Z
\gamma$ and $ZZZ$ vertices. We discuss, in a systematic way, the
tensor structure of the the three-point vertex. Starting with the most
general CP conserving vertex, we first use Schouten's identity to
reduce the number of independent form-factors. Gauge invariance and
Bose symmetry, whenever applicable, are used then to obtain the
required structure. We have then provided explicit calculations for
the form-factors~\cite{Gaemers,Hagiwara} relevant for these vertices
at different limits, particularly when two out of these gauge bosons
in a vertex are on-shell. We define total form-factors, ${\cal
H}^{\gamma/Z}_{3,4}$ and ${\cal F}^{\gamma/Z}_5$, which are found to
be more appropriate than the conventional (${h}^{\gamma/Z}_{3,4}$ and
${f}^{\gamma/Z}_5$) ones in discussions of off-shell behaviour.

At the one-loop level, only fermions contribute to these form-factors.
Within the SM, these are the quarks and leptons, while, in the context
of the MSSM, the charginos contribute too (neutralinos come into play
only as far the $ZZZ$ vertex is concerned).  We have provided explicit
formulae for the generic form-factor as obtained from such loop
diagrams. The calculations clearly demonstrate that ${\cal
H}^{\gamma/Z}_{4}=0$ to this order, a conclusion in agreement with the
observation made by Gounaris {\it et al.}~\cite{gounaris2}.

 We have studied the $Q^2$ dependence of these form-factors when two of
the gauge bosons are on-shell for both SM and MSSM. Here we would like
to emphasize the fact that these three point vertices can take part in
$t$-channel processes and keeping this in mind we have presented
values for both positive and negative $Q^2$. Cusps and peaks appear at
the different thresholds defined by the masses of the internal
fermions. The maximum magnitudes for the form-factors can be as high
as $|Re[{\cal H}^{\gamma}_{3}]|\simeq 9\times 10^{-3}$ for SM at the
top threshold, whereas for MSSM, the magnitude is smaller in most
cases, the most promising one being $|Im[{\cal F}^{Z}_5]|\simeq
9\times 10^{-4}$ which depends on chargino and neutralino masses
which, in turn, are parameter space dependent. One can expect this
enhancement for ${\cal F}^{Z}_5$ due to almost degeneracy of the
lightest chargino and next to lightest neutralino characteristic of
this particular space and consequent opening of thresholds at similar
$Q^2$ values.

These new physics effects are model dependent. So we have studied next
the MSSM parameter space dependence of these form-factors. The contour
plots of Fig.~5 should turn out to be {\em useful to exclude certain
regions of the parameter space} if such new physics effects are
estimated experimentally. 

To get a better hold in determining these new physics effects, we have
studied the effects on these form-factors when all the bosons are
off-shell. It might turn out to be {\em handy in identifying these
effects} as we can then have a better feeling of maximizing SUSY
contributions playing with the off-shell vector bosons.

In short we have not only reviewed the SM contribution to triple
neutral gauge boson vertices which will be relevant for a better
understanding of the non-abelian structure of the SM, but also studied
in detail the possible MSSM contributions to the same.  These
numerical estimations not only provide an independent verification of
results provided by Ref.~\cite{gounaris2} but {\em complement their
results} by providing the negative $Q^2$ values.  Our study might also
turn out to be useful to disentangle MSSM effects from the SM ones.

\vskip 20pt
\noindent
{\bf Acknowledgements}\\ We would like to thank WHEPP5
organisers~\cite{whepp} where this work was initiated. S. Rakshit
acknowledges partial support from the Council of Scientific and
Industrial Research, India.  Both S. Rakshit and S. Dutta would like
to thank the Mehta Research Institute, Allahabad, for their
hospitality while part of the work was being carried out.

\end{document}